\documentclass[%
 reprint,
 amsmath,amssymb,
 aps,
]{revtex4-2}

\usepackage{graphicx}
\usepackage{dcolumn}
\usepackage{bm}
\usepackage{xcolor}
\usepackage{textcomp}
\usepackage{float}

\usepackage{braket} 
\usepackage{caption}
\usepackage{subcaption}
\usepackage{booktabs}
\usepackage{algorithm}
\usepackage{multirow}
\usepackage[noend]{algpseudocode}
\usepackage{tikz}
\usetikzlibrary{shapes.geometric, arrows.meta, positioning, calc}
\usetikzlibrary{arrows.meta,positioning,calc}
\everymath{\displaystyle}

\DeclareCaptionFormat{myformat}{#3}
\begin{document}

\preprint{APS/123-QED}


\title{Physics-informed neural networks for viscoelastic fluid
flows around a cylinder in a two-dimensional  channel
}

\author{Midhuna Suresh}
 \email{midhuna000suresh@gmail.com}
\affiliation{Department of Physics, Maulana Azad National Institute of Technology (MANIT), Bhopal-462003, India}
\author{Akanksha Gupta}
\email{akanksha@manit.ac.in}
\affiliation{Department of Physics, Maulana Azad National Institute of Technology (MANIT), Bhopal-462003, India}

\date{\today}

\begin{abstract}
Grid-based fluid dynamics solvers routinely struggle with exhaustive meshing demands and ill-posed inverse problems. A practical mesh-free alternative is given by Physics-informed neural networks (PINNs) but, applying them to highly elastic Oldroyd-B fluids results in many training failures.  The High Weissenberg Number Problem (HWNP), driven by the exponential stress growth near stagnation points is a major issue, which causes standard PINN optimizers to diverge. To prevent the network from crashing, we apply a Cholesky decomposition to the conformation tensor. This mathematical constraint stabilizes the gradients by guaranteeing a positive-definite stress field. Beyond mathematical stability, the inherent spectral bias of deep learning models can hinder the network from accurately capturing the highly elastic wake structures. Therefore, we used sparse data assimilation to force the model toward the actual physical solution. By anchoring the physics loss with targeted CFD data points and accelerating training via transfer learning, we successfully pushed the network past non-physical local minima. We validated this Cholesky-PINN approach on flow past cylindrical geometries for Reynolds numbers ($Re$) between 5-25, for the single-cylinder setup. In addition, the relaxation time ($\lambda$) is increased from 0.1 to 0.5 to test the stability of the network. Finally, we scale the framework to a complex 3-cylinder array which proved our constructive solid geometry approach completely bypasses the tedious re-meshing steps of traditional CFD. The combined framework accurately captured sharp viscoelastic wakes, providing a stable computational tool for complex rheological modeling.
\end{abstract}

\keywords{Physics-informed neural networks (PINNs), Viscoelastic Fluids, Oldroyd-B}
\maketitle


\noindent 

\section{Introduction}

 In continuum mechanics, the flow of a fluid past a cylindrical obstacle is a canonical, rigorously studied benchmark~\cite{alves2001flow, kuzmina2021flow, wang2013numerical,gajendragad2026} where, bluff bodies like cylinders induce sharp adverse pressure gradients, unlike streamlined shapes. This makes them a perfect isolated environment to study complex phenomena like boundary layer separation, wake elongation, and vortex shedding \cite{batchelor2000introduction, kundu2024fluid}. The circular cylinder acts as a foundational model because it breaks down these complicated fluid mechanics into a simplified geometric framework. Mastering these isolated kinematics before scaling up to complex engineering designs is needed, whether they are microfluidic devices, massive aerospace structures, or tandem multi-body arrays \cite{zhou2020viscoelastic, singha2016numerical}.

Traditional Computational Fluid Dynamics (CFD)~\cite{baaijens1998mixed, alves2021numerical, schneiderbauer2014navier,verma_2020,Akanksha_PhysRevE.100.053101,Aka4th, Gupta_POP2014,Mukherjee_2019} has long been the standard tool required to resolve the foundational Navier-Stokes momentum equations \cite{navier1838navier, bistafa2018development, lukaszewicz2016navier}. Even though grid-based solvers using finite volume or finite element methods are highly accurate for well-posed forward simulations, they suffer from severe operational bottlenecks. Classical CFD pipelines demand exhaustive, geometry-specific mesh generation \cite{zhao2024comprehensive, castillo2022numerical}. For complex or dynamically moving domains, this requirement quickly becomes computationally prohibitive. Conventional solvers are even more problematic in handling ill-posed inverse problems, such that conventional CFD simply cannot bridge the gap if a fluid system has unknown parameters, noisy boundaries, or relies on scattered observational data~\cite{karniadakis2021physics}.

These mathematical limitations drove the rapid adoption of scientific machine learning. Physics-Informed Neural Networks (PINNs) are at the forefront of this shift~\cite{raissi2019physics}, rapidly evolving through advanced spatial and domain-decomposition techniques to handle increasingly complex topologies ~\cite{Jagtap2020_XPINN, Jagtap2020_cPINN, Kharazmi2020_hpVPINN, Sirignano2018_DGM,singh2026physicsinformedkolmogorovarnoldnetworksviscoelastic}. By embedding governing partial differential equations directly into a deep learning loss function,  PINNs act as physically constrained surrogate models and this allows the network to assimilate scattered, multi-fidelity data to infer latent physical quantities. It identifies unmodeled physics directly from the data, without requiring a spatial mesh~\cite{jin2021nsfnets, cai2021flow}.  Yet, standard PINN architectures still hide computational vulnerabilities despite their success in such data-scarce scenarios. A major hurdle is the spectral bias often called the ``F-principle'' \cite{Rahaman2019SpectralBias}. While fully connected neural networks naturally tend to converge quickly on low-frequency, smooth functions, they fail to capture sharp, high-frequency gradients. This means standard PINNs often smear or completely miss abrupt flow separations and steep wake profiles for multiscale fluid dynamics. The optimizer easily gets trapped in non-physical local minima because the different loss components compete during gradient descent \cite{Krishnapriyan2021FailureModes, wang2021understanding}. This creates the necessity of integrating data assimilation techniques to beat this spectral biasing and stop the training stagnation in highly non-linear flow regions and anchoring the network with sparse, high-fidelity observational points aggressively guides the optimizer toward the correct physical solution. This strategy has proven highly effective in resolving complex mean-flow reconstructions and turbulent regimes \cite{Raissi2018_HFM_arXiv, hanrahan2023studying, patel2024turbulence, jang2024physics, ghosh2024geometry}. 

These computational flaws become critical when moving from Newtonian fluids to highly elastic non-Newtonian regimes~\cite{Larson1999Constitutive} howeverthis gap has been bridged by extending the foundational mesh-free framework established by Ang \textit{et al.}~\cite{ang2023physics} into the highly unstable viscoelastic regime governed by the Oldroyd-B constitutive model \cite{Oldroyd1950, SHAQFEH2021104672, Bird1987, article, 10.1093/qjmam/13.4.444}. Interestingly, these viscoelastic constitutive formulations extend far beyond polymer melts. They accurately describe dynamics strongly coupled dusty plasmas in fluid regime using generalized hydrodynamic model~\cite{Aka4th, Aka, aka6th}.

Usually, the separation zones and elongated wakes behind cylindrical obstacles are dominated by strong extensional kinematics; however, here localized extensional strain forces the polymeric stress tensor to grow exponentially. As the flow approaches critical Weissenberg numbers, the growing polymeric stress drives exceptionally sharp spatial and velocity gradients \cite{thompson2021reynolds, lauga2009life, afonso2011dynamics, kumar2023lagrangian, hopkins2022upstream, kumar2022hysteresis}.  Traditional grid-based methods cannot handle these unbounded gradients. The High Weissenberg Number Problem (HWNP) \cite{review2022, sanchez2022understanding, chokshi2007stability} occurs when the solver diverges immediately as a result of the conformation tensor losing its Symmetric Positive-Definite (SPD) property. Standard PINN architectures suffer the same fate; the HWNP induces pathological partial differential equation residuals capable of trapping unconstrained networks in non-physical states \cite{mahmoudabadbozchelou2022nn, sa2025physics}.

To overcome these stability limits and unconditionally guarantee a valid stress field, we reformulated the neural network architecture by integrating a Cholesky decomposition directly into the loss function \cite{pires2023stabilization}. Factorizing the conformation tensor as $\mathbf{c} = \mathbf{L}\mathbf{L}^T$ forces the predicted stress field to remain strictly SPD at every spatial coordinate, offering a robust alternative to matrix-logarithm reconstructions \cite{hulsen2005flow, fattal2005time, afonso2009log} and immunizing the optimizer against gradient explosions. Mathematical constraints stabilize the network, however, they do not solve spectral bias on their own. Hybrid data-assimilation was mandatory in our work to capture the sharp, high-frequency spatial gradients inside viscoelastic wakes. The network was pulled out of unphysical local minima by anchoring the physics-loss with a sparse subset of high-fidelity coordinate points. This allows the model to accurately resolve extreme viscoelastic kinematics.

The primary objective of this study is to establish a stabilized, data-driven deep learning framework. It must forward-simulate highly viscoelastic flows without the stability limits of traditional CFD or the training stagnation of standard PINNs. The specific contributions of this work are threefold:

\begin{itemize}
    \item Developing a Cholesky-PINN architecture preserving the SPD conformation tensor to resolve the HWNP in Oldroyd-B fluids.
    \item Implementing targeted sparse data assimilation and transfer learning to overcome spectral bias at elevated elasticity and inertial limits ($Re = 5$ to $25$) \cite{wang2025transfer}.
    \item Scaling this mesh-free framework to a complex, staggered three-cylinder array to prove Constructive Solid Geometry is superior to traditional grid-based re-meshing \cite{lu2021deepxde}.
\end{itemize}

\section{Methodology}

\subsection{Problem Setup and Computational Domain}

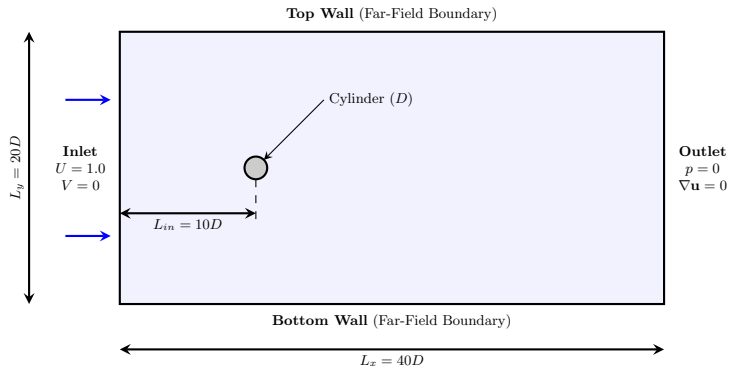
\begin{figure}[htbp]
\centering
\begin{tikzpicture}[x=1cm, y=1cm, >=stealth, scale=0.6, transform shape]
    \def\Lx{12}
    \def\Ly{6}
    \def\Cx{3} 
    \def\Cy{3} 
    \def\R{0.25} 

    \draw[thick, fill=blue!5] (0,0) rectangle (\Lx,\Ly);

    \draw[thick, fill=gray!40] (\Cx,\Cy) circle (\R);

    \node[left, align=center, font=\small, xshift=-0.2cm] at (0, \Ly/2) {\textbf{Inlet}\\$U = 1.0$\\$V = 0$};
    \node[right, align=center, font=\small, xshift=0.2cm] at (\Lx, \Ly/2) {\textbf{Outlet}\\$p = 0$\\$\nabla \mathbf{u} = 0$};
    \node[above, font=\small, yshift=0.1cm] at (\Lx/2, \Ly) {\textbf{Top Wall} (Far-Field Boundary)};
    \node[below, font=\small, yshift=-0.1cm] at (\Lx/2, 0) {\textbf{Bottom Wall} (Far-Field Boundary)};

    \draw[<->, thick] (0,-1.0) -- (\Lx,-1.0) node[midway, below] {$L_x = 40D$};
    \draw[<->, thick] (-2.0,0) -- (-2.0,\Ly) node[midway, above, rotate=90] {$L_y = 20D$};
    
    \draw[<->, thick] (0, \Cy-1.0) -- (\Cx, \Cy-1.0) node[midway, below, font=\small] {$L_{in} = 10D$};
    \draw[dashed, thin] (\Cx, \Cy-\R) -- (\Cx, \Cy-1.2); 
    
    \foreach \y in {1.5, 4.5} {
        \draw[->, thick, blue] (-1.2, \y) -- (-0.2, \y);
    }
    
    \draw[<-] (\Cx+\R*0.7, \Cy+\R*0.7) -- (\Cx+1.5, \Cy+1.5) node[right, font=\small] {Cylinder ($D$)};
\end{tikzpicture}
\caption{Schematic of the computational domain and boundary conditions for the single confined cylinder.}
\label{fig:domain_single}
\end{figure}

We establish the computational domain, as seen in Fig.~\ref{fig:domain_single}, to simulate the viscoelastic flow past cylindrical obstacles. The primary configuration consists of a single circular cylinder with a diameter $D = 0.1$ m. To isolate the localized wake dynamics from artificial boundary reflections, we place the obstacle asymmetrically inside a large rectangular channel. The total domain spans a length of $L_x = 40D$ and a height of $L_y = 20D$. We position the center of the cylinder at the coordinates $(10D, 10D)$. 

At the left inlet boundary, the fluid enters with a uniform velocity profile ($U = 1.0$ m/s, $V = 0$ m/s) and an unstretched, equilibrium polymer state ($l_{11} = 1.0$, $l_{22} = 1.0$, $l_{21} = 0.0$). The standard no-slip condition ($\mathbf{u} = 0$) applies for the solid cylinder walls. The downstream right boundary acts as an outlet, constrained by a zero reference pressure ($p = 0$) and a Neumann zero-gradient velocity condition ($\nabla \mathbf{u} = 0$). 

\begin{figure}[htbp]
\centering
\begin{tikzpicture}[x=1cm, y=1cm, >=stealth, scale=0.5, transform shape]
    \def\Lx{12}
    \def\Ly{6}
    \def\R{0.3} 
    
    \def\ConeX{3.0}   \def\ConeY{3.0}
    \def\CtwoX{3.8}   \def\CtwoY{3.5}
    \def\CthreeX{3.8} \def\CthreeY{2.5}

    \draw[thick, fill=blue!5] (0,0) rectangle (\Lx,\Ly);

    \draw[thick, fill=gray!40] (\ConeX,\ConeY) circle (\R);
    \draw[thick, fill=gray!40] (\CtwoX,\CtwoY) circle (\R);
    \draw[thick, fill=gray!40] (\CthreeX,\CthreeY) circle (\R);

    \draw[dashed, red, thick] (\ConeX,\ConeY) -- (\CtwoX,\CtwoY) -- (\CthreeX,\CthreeY) -- cycle;

    \node[left, align=center, font=\small] at (-0.2, \Ly/2) {\textbf{Inlet}\\$U = 1.0$\\$V = 0$};
    \node[right, align=center, font=\small] at (\Lx+0.2, \Ly/2) {\textbf{Outlet}\\$p = 0$\\$\nabla \mathbf{u} = 0$};
    \node[above, font=\small] at (\Lx/2, \Ly+0.1) {\textbf{Top Wall} (Far-Field Boundary)};
    \node[below, font=\small] at (\Lx/2, -0.1) {\textbf{Bottom Wall} (Far-Field Boundary)};

    \draw[<->, thick] (0,-1.2) -- (\Lx,-1.2) node[midway, below] {$L_x = 40D$};
    \draw[<->, thick] (-2.2,0) -- (-2.2,\Ly) node[midway, above, rotate=90] {$L_y = 20D$};
    
    \draw[<->, thick] (0, \ConeY) -- (\ConeX-\R, \ConeY) node[midway, above, font=\small] {$L_{in} = 10D$};
    
    \foreach \y in {1.5, 4.5} {
        \draw[->, thick, blue] (-1.5, \y) -- (-0.5, \y);
    }
    
    \draw[<-] (\CtwoX+\R, \CtwoY+\R) -- (\CtwoX+1.5, \CtwoY+1.5) node[right, align=left, font=\small] {Staggered Array\\(Gap $\approx 0.5D$)};
\end{tikzpicture}
\caption{2D flow domain and boundary constraints detailing the three-cylinder array.}
\label{fig:domain_three}
\end{figure}
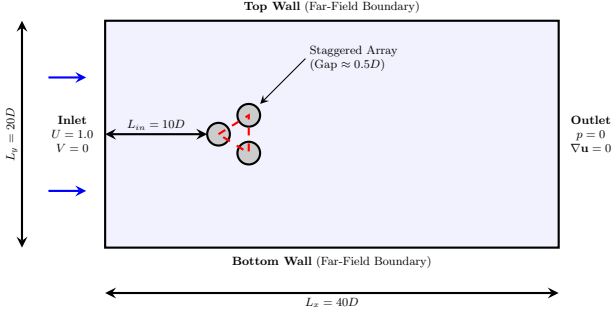

To thoroughly stress-test the framework, we expanded the geometry into a staggered three-cylinder array as shown in Fig \ref{fig:domain_three} \cite{singha2016numerical, kumar2022hysteresis}. While the outer domain boundaries remain unchanged, the internal setup features a leading cylinder at $(10D, 10D)$ followed by two trailing cylinders. These rear cylinders are positioned downstream at a longitudinal coordinate of $11.5D$ and shifted symmetrically along the transverse axis to $10.86D$ and $9.14D$. Arranging them in this tight equilateral formation forces the fluid through an exceptionally narrow interstitial gap of roughly 0.5D between the solid surfaces.

\subsection{Domain Discretization and Data Generation}

Extracting our baseline data required a traditional finite volume approach, so we ran our simulations in the open-source framework OpenFOAM \cite{kornbleuth2016studying}. We specifically relied on the \textit{rheoTool} toolbox because of its ability to handle complex viscoelastic flow fields \cite{mohanty2024influence}. While traditional discrete solvers demand extremely refined grids just to handle narrow gaps—often maxing out computational memory limits \cite{castillo2022numerical, zhao2024comprehensive}—our neural network avoids grid generation entirely \cite{lu2021deepxde}. Inside DeepXDE, the main fluid channel is defined as a rectangular primitive and the solid cylinders are carved out using Constructive Solid Geometry (CSG) boolean operations.

Instead of building a connected mesh, we populate the fluid domain by scattering 20,000 internal collocation points via Latin Hypercube Sampling (LHS). 2,000 explicit points along the boundaries are also assigned. Because these scattered coordinates lack any topological connectivity, there is zero mesh overhead. They exist purely as spatial nodes where automatic differentiation evaluates the exact analytical gradients for both the Navier-Stokes equations and the Cholesky-conformation laws.

\subsection{Mathematical Formulation}

\subsubsection{Governing Conservation Laws}
This study models a viscoelastic fluid inside a strictly two-dimensional domain. By assuming the flow remains steady, isothermal, and incompressible, the standard mass and momentum conservation laws \cite{lukaszewicz2016navier, batchelor2000introduction, kundu2024fluid} reduce to the following expressions:
$$ \nabla \cdot \mathbf{u} = 0 $$
$$ \rho (\mathbf{u} \cdot \nabla)\mathbf{u} = -\nabla p + \nabla \cdot \boldsymbol{\tau} $$
Here, $\mathbf{u}$ is the velocity vector field, $p$ is the thermodynamic pressure, $\rho$ is the fluid density, and $\boldsymbol{\tau}$ represents the total extra-stress tensor. Since we restrict our analysis to low Reynolds number regimes ($Re = 5-25$), the flow remains strictly laminar.

\subsubsection{The Oldroyd-B Constitutive Model}

To close the momentum equation, we must define the fluid's rheological behavior. We utilize the Oldroyd-B constitutive model to simulate the creeping flow of the viscoelastic fluid over the channel-confined circular cylinders \cite{Oldroyd1950, Renardy2021, mohanty2024influence, Bird1987, Larson1999Constitutive}. This framework splits the total extra-stress tensor into two distinct parts: a Newtonian solvent contribution ($\boldsymbol{\tau}_s$) and an elastic polymeric contribution ($\boldsymbol{\tau}_p$).
$$ \boldsymbol{\tau} = \boldsymbol{\tau}_s + \boldsymbol{\tau}_p $$
The solvent stress follows a standard linear Newtonian relationship based on the solvent viscosity $\eta_s$.
$$ \boldsymbol{\tau}_s = \eta_s (\nabla \mathbf{u} + (\nabla \mathbf{u})^T) $$
The polymeric stress introduces the fluid's fading memory and elastic recoil \cite{boger1977highly}. It relies on a complex differential transport equation. Instead of computing the stress directly, it is mathematically safer to track the microstructural deformation of the polymer chains. We represent this stretching using the dimensionless conformation tensor $\mathbf{c}$. The polymeric stress relates to this tensor through the polymer viscosity $\eta_p$ and the relaxation time $\lambda$:
$$ \boldsymbol{\tau}_p = \frac{\eta_p}{\lambda} (\mathbf{c} - \mathbf{I}) $$
where $\mathbf{I}$ is the identity tensor. The conformation tensor itself evolves according to the upper-convected material derivative, which balances the elastic relaxation with the convective flow deformation:
$$ \mathbf{c} + \lambda \left( (\mathbf{u} \cdot \nabla)\mathbf{c} - (\nabla \mathbf{u})\mathbf{c} - \mathbf{c}(\nabla \mathbf{u})^T \right) = \mathbf{I} $$

\subsubsection{Cholesky Decomposition of the Conformation Tensor}
The standard Oldroyd-B formulation breaks down at high Weissenberg numbers \cite{alves2021numerical}. Extensional kinematics near the cylinder stagnation points stretch the polymer chains severely. This forces the conformation tensor to grow exponentially. Traditional grid-based solvers cannot handle these steep gradients. Numerical diffusion eventually causes the conformation tensor to lose its Symmetric Positive-Definite (SPD) property \cite{afonso2009log, hulsen2005flow, fattal2005time}. Once the matrix yields negative eigenvalues, the entire solver crashes.

We reformulate the conformation tensor using a Cholesky decomposition to guarantee unconditional physical validity inside the neural network. We express the tensor as the product of a lower-triangular matrix $\mathbf{L}$ and its transpose.
$$ \mathbf{c} = \mathbf{L} \mathbf{L}^T $$
In a two-dimensional framework, the matrix $\mathbf{L}$ contains three independent components:
$$ \mathbf{L} = \begin{bmatrix} l_{11} & 0 \\ l_{21} & l_{22} \end{bmatrix} $$
The neural network directly predicts these $l_{ij}$ components instead of the raw stress values. We algebraically reconstruct the conformation tensor before calculating the physics residuals. Because any matrix multiplied by its transpose is strictly positive-definite, this mathematical constraint permanently immunizes the optimizer against the High Weissenberg Number Problem. The network can never predict a physically impossible negative stress state.

\subsection{Physics-Informed Neural Network Architecture}

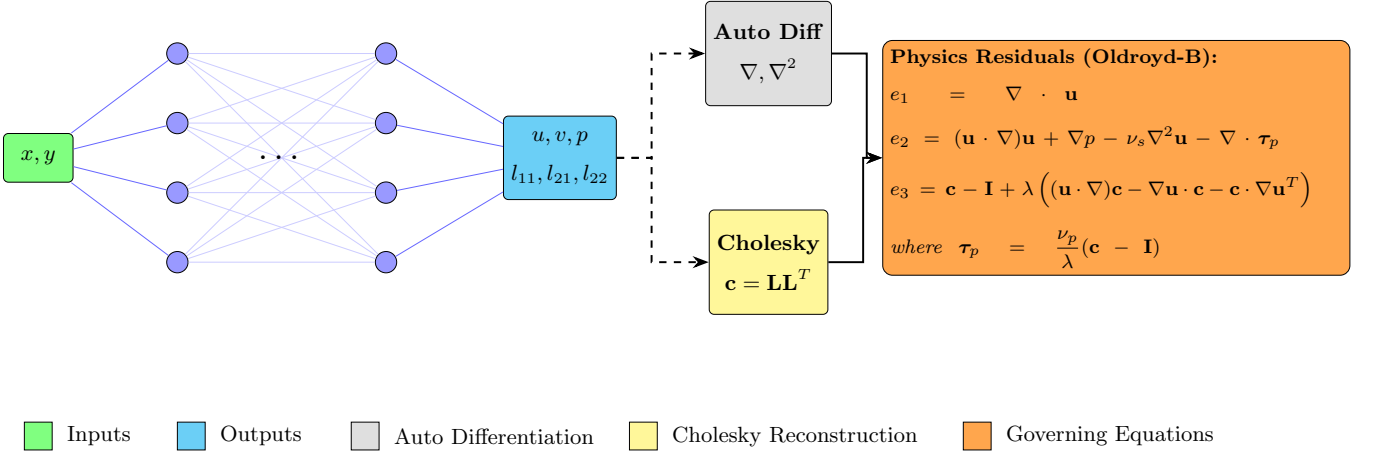
\begin{figure*}[htbp]
\centering
\resizebox{1.0\textwidth}{!}{%
\begin{tikzpicture}[>=Stealth]

\node[draw, fill=green!50, minimum width=1cm, minimum height=0.7cm, rounded corners=2pt]
(in) at (0,0) {$x,y$};

\node[circle, draw, fill=blue!40] (h11) at (2, 1.5) {};
\node[circle, draw, fill=blue!40] (h12) at (2, 0.5) {};
\node[circle, draw, fill=blue!40] (h13) at (2, -0.5) {};
\node[circle, draw, fill=blue!40] (h14) at (2, -1.5) {};

\node[circle, draw, fill=blue!40] (h21) at (5, 1.5) {};
\node[circle, draw, fill=blue!40] (h22) at (5, 0.5) {};
\node[circle, draw, fill=blue!40] (h23) at (5, -0.5) {};
\node[circle, draw, fill=blue!40] (h24) at (5, -1.5) {};

\node[draw, fill=cyan!50, minimum width=1.5cm, minimum height=1.2cm, rounded corners=2pt, align=center]
(out) at (7.5,0) {$u, v, p$ \\[2mm] $l_{11}, l_{21}, l_{22}$};

\foreach \i in {h11,h12,h13,h14}
    \draw[blue!60] (in) -- (\i);

\foreach \i in {h11,h12,h13,h14}
{
  \foreach \j in {h21,h22,h23,h24}
  {
    \draw[blue!20] (\i) -- (\j);
  }
}

\foreach \i in {h21,h22,h23,h24}
    \draw[blue!60] (\i) -- (out);

\node at (3.5,0) {\Large $\cdots$};

\node[
draw,
fill=gray!25,
minimum width=1.2cm,
minimum height=1.5cm,
align=center,
rounded corners=2pt
] (ad) at (10.5, 1.5)
{
\textbf{Auto Diff}\\[2mm]
$\nabla, \nabla^2$
};

\node[
draw,
fill=yellow!50,
minimum width=1.2cm,
minimum height=1.5cm,
align=center,
rounded corners=2pt
] (chol) at (10.5, -1.5)
{
\textbf{Cholesky}\\[2mm]
$\mathbf{c} = \mathbf{L}\mathbf{L}^T$
};

\draw[dashed, ->, thick] (out.east) -- ++(0.5,0) |- (ad.west);
\draw[dashed, ->, thick] (out.east) -- ++(0.5,0) |- (chol.west);

\node[
draw,
rounded corners=4pt,
fill=orange!70,
text width=6.5cm,
align=left
] (eq) at (15.5, 0)
{
\footnotesize
\textbf{Physics Residuals (Oldroyd-B):}\\[2mm]
$e_1 = \nabla \cdot \mathbf{u}$\\[3mm]
$e_2 = (\mathbf{u} \cdot \nabla)\mathbf{u} + \nabla p - \nu_s \nabla^2 \mathbf{u} - \nabla \cdot \boldsymbol{\tau}_p$\\[3mm]
$e_3 = \mathbf{c} - \mathbf{I} + \lambda \left( (\mathbf{u} \cdot \nabla)\mathbf{c} - \nabla\mathbf{u} \cdot \mathbf{c} - \mathbf{c} \cdot \nabla\mathbf{u}^T \right)$\\[3mm]
\textit{where } $\boldsymbol{\tau}_p = \frac{\nu_p}{\lambda} (\mathbf{c} - \mathbf{I})$
};

\draw[->, thick] (ad.east) -- ++(0.5,0) |- (eq.west);
\draw[->, thick] (chol.east) -- ++(0.5,0) |- (eq.west);

\node[draw, fill=green!50, minimum size=4mm] at (0, -4) {};
\node[right=1mm] at (0.2, -4) {Inputs};

\node[draw, fill=cyan!50, minimum size=4mm] at (2.2, -4) {};
\node[right=1mm] at (2.4, -4) {Outputs};

\node[draw, fill=gray!25, minimum size=4mm] at (4.7, -4) {};
\node[right=1mm] at (4.9, -4) {Auto Differentiation};

\node[draw, fill=yellow!50, minimum size=4mm] at (8.7, -4) {};
\node[right=1mm] at (8.9, -4) {Cholesky Reconstruction};

\node[draw, fill=orange!70, minimum size=4mm] at (13.5, -4) {};
\node[right=1mm] at (13.7, -4) {Governing Equations};

\end{tikzpicture}
}
\caption{Schematic representation of the Cholesky-PINN architecture for solving the 2D incompressible Oldroyd-B equations. The network outputs kinematic fields and lower-triangular components ($l_{ij}$), which undergo Cholesky reconstruction to strictly enforce a Symmetric Positive-Definite conformation tensor ($\mathbf{c}$) prior to residual evaluation.}
\label{fig:cholesky_pinn_schematic}
\end{figure*}

We construct a fully connected deep neural network using the DeepXDE library \cite{lu2021deepxde, karniadakis2021physics}, integrating specialized architectures designed for complex fluid modeling and stress discovery \cite{mahmoudabadbozchelou2022nn, thakur2024viscoelasticnet}. The architecture accepts spatial coordinates $(x,y)$ as inputs. It maps these coordinates to the primitive kinematic fields $(u, v, p)$ alongside the three lower-triangular components of the conformation tensor ($l_{11}, l_{21}, l_{22}$). Fig.\ref{fig:cholesky_pinn_schematic} illustrates the complete forward-pass workflow. We enforce the Symmetric Positive-Definite constraint algebraically before the automatic differentiation engine calculates the spatial gradients \cite{pires2023stabilization, sa2025physics}. This strictly stabilizes the non-linear transport terms inside the Oldroyd-B residuals and protects the optimizer from diverging \cite{sanchez2022understanding}.

To navigate this highly complex optimization landscape, which is prone to numerous local minima, we implement a sequential two-stage optimization strategy \cite{DeRyck2024_PINN_analysis, Cui_POF2025}. The network initially trains utilizing the Adam optimizer. This first-order algorithm rapidly navigates the broad parameter space and prevents the model from stagnating in shallow error valleys \cite{wang2021understanding}. Once the loss trajectory stabilizes, we transition the network to the quasi-Newton Limited-memory Broyden-Fletcher-Goldfarb-Shanno (L-BFGS) optimizer \cite{lu2021deepxde}. This second-order method exploits local gradient curvature to aggressively drive the physical and boundary residuals down to machine precision.

\subsection{Data Assimilation and Transfer Learning}
Purely physics-driven networks frequently fail when modeling highly elastic flows. Deep networks naturally prioritize low-frequency functions and ignore sharp stress gradients \cite{Rahaman2019SpectralBias, Krishnapriyan2021FailureModes}. This spectral bias traps the optimizer in non-physical solutions \cite{wang2021understanding}. To force the model to capture the complex wake signatures, we embed a sparse data assimilation strategy \cite{Raissi2018_HFM_arXiv, cai2021flow, dou2025flow}. We extract a highly restricted subset of spatial coordinates from the converged OpenFOAM baseline simulations \cite{kornbleuth2016studying, zolman2025sindy}. By anchoring the physics-loss with these targeted physical points, we aggressively pull the neural network out of stagnant local minima \cite{zhao2024comprehensive, mangal2025learning}. The composite loss function minimizes the boundary conditions, the physical PDE residuals, and the assimilated data points simultaneously. We assign a heavy penalty weight to the data anchors to guarantee strict physical compliance across the steep elastic boundary layers.

Resolving the coupled momentum and conformation equations demands rigorous hyperparameter tuning and optimization \cite{Cui_POF2025, eshkofti2024modified}. We employ a sequential, two-stage optimization protocol \cite{karniadakis2021physics}. The Adam optimizer first navigates the broad loss landscape. The quasi-Newton L-BFGS algorithm then refines the predictions down to machine precision. Training a network from a random initialization becomes computationally prohibitive at higher Reynolds numbers or extreme elasticities. We implement transfer learning to bypass this bottleneck \cite{wang2025transfer, cui2026coupled}. We reuse the converged weights from a low-inertia baseline simulation as the starting point for a more complex flow regime. This initialization provides an incredibly accurate starting distribution, allowing us to safely truncate the Adam optimization phase and drastically reduce overall computational overhead.

Table \ref{tab:oldroyd_parameters} details the baseline thermophysical parameters defining the Oldroyd-B simulations \cite{mohanty2024influence, sasmal2020combined, thompson2021reynolds}. We maintain a strict viscosity ratio of $\beta = 0.5$ between the Newtonian solvent and the elastic polymer phases to replicate standard highly elastic constant-viscosity Boger fluids \cite{boger1977highly, SHAQFEH2021104672}.

\begin{table*}[htbp]
\centering
\renewcommand{\arraystretch}{1.3} 
\begin{tabular}{c l c}
\hline
\textbf{Symbol} & \textbf{Description} & \textbf{Value } \\
\hline
$\rho$ & Fluid density & $1000 \ \mathrm{kg/m^3}$ \\
$\eta$ & Total dynamic viscosity ($\rho \nu$) & $20 \ \mathrm{Pa\cdot s}$ \\
$\eta_s$ & Solvent viscosity ($\beta \eta$) & $10 \ \mathrm{Pa\cdot s}$ \\
$\eta_p$ & Polymer viscosity ($(1-\beta)\eta$) & $10 \ \mathrm{Pa\cdot s}$ \\
$\lambda$ & Polymer relaxation time & $0.1, 0.3, 0.5 \ \mathrm{s}$ \\
$U$ & Characteristic velocity scale (Inlet) & $1.0 \ \mathrm{m/s}$ \\
$D$ & Characteristic length scale (Cylinder) & $0.1 \ \mathrm{m}$ \\
$\mathrm{Re}$ & Reynolds number ($\rho U D / \eta$) & $5, 10, 15, 20, 25$ \\
$\mathrm{Wi}$ & Weissenberg number ($\lambda U / D$) & $1.0, 3.0, 5.0$ \\
\hline
\end{tabular}
\caption{Thermophysical parameters and dimensionless groups defining the Oldroyd-B simulations.}
\label{tab:oldroyd_parameters}
\end{table*}

\section{Results and Discussion}

\subsection{Pure PINN at $Re=5$, $\lambda=0.1$}

\begin{figure*}[htbp]
    \centering
    \includegraphics[width=0.85\textwidth]{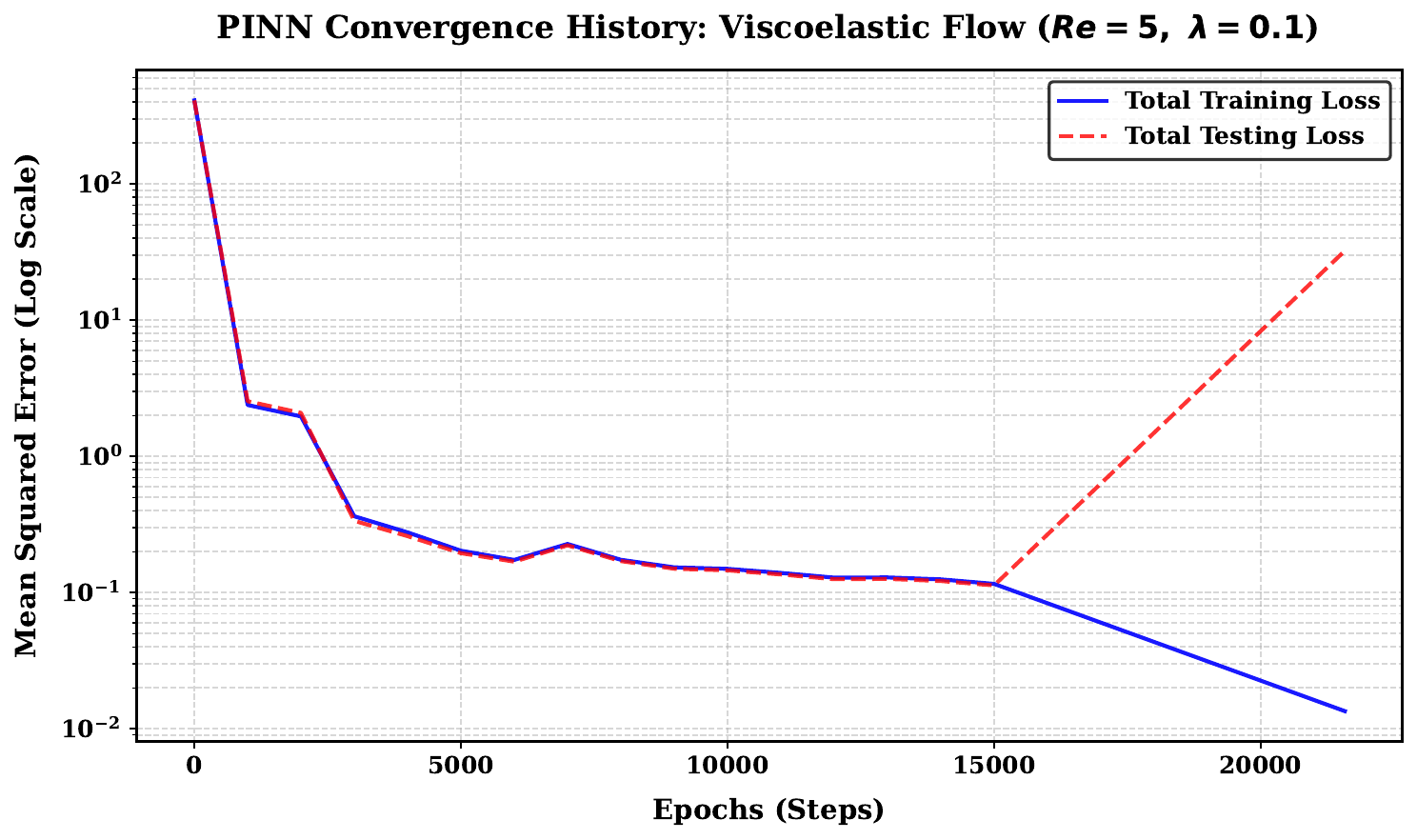}
    \caption{Convergence history for the pure physics-informed neural network at $Re = 5$ ($\lambda = 0.1$).}
    \label{fig:Re5_Loss}
\end{figure*}

\begin{figure*}[htbp]
    \centering
    \includegraphics[width=1.08\textwidth]{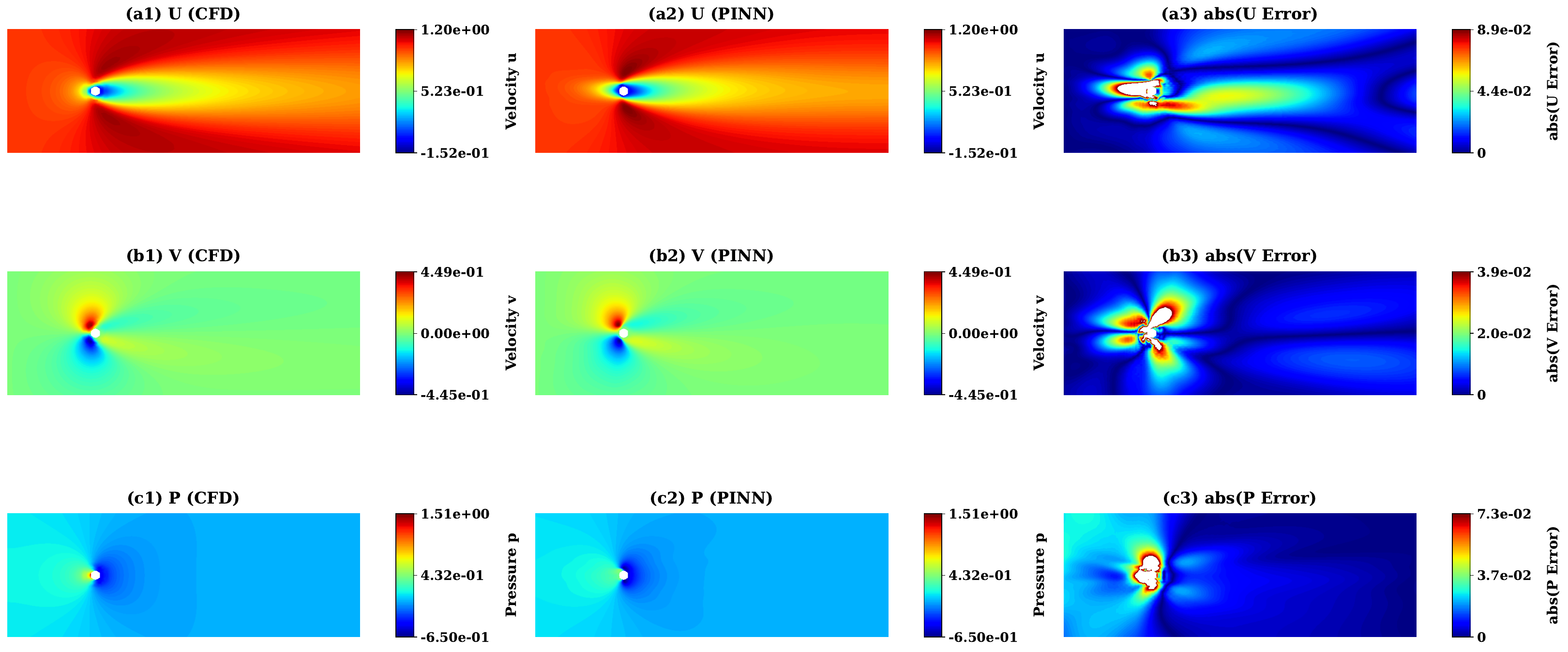}
    \caption{For $\lambda = 0.1$ and $Re = 5$, a side-by-side comparison of the flow fields.}
    \label{fig:Re=5_cfd_vs_PINNs_comparison_master}
\end{figure*}

We started by training the network entirely from scratch on the baseline creeping flow ($Re=5$, $\lambda=0.1$). By relying solely on physical boundaries the spatial coordinates $(x, y)$ were mapped directly to the primary solution vector $[u, v, p, l_{11}, l_{21}, l_{22}]$. As shown in Fig.\ref{fig:Re5_Loss}, this purely physical solver initially converges, howeverthe testing loss (red dashed line) sharply diverges just after epoch 15,000. This is a classic case of collocation point overfitting. Because the Oldroyd-B equations are severely stiff, the quasi-Newton L-BFGS optimizer essentially memorized the exact training coordinates, leading to a complete failure in the interstitial spaces. Fig.\ref{fig:Re=5_cfd_vs_PINNs_comparison_master} perfectly captures this breakdown, showing localized absolute errors peaking at $8.9\text{e-}02$ for the velocity field u. This failure is a critical finding. It provides the definitive mathematical justification for why purely physical PINNs are insufficient for viscoelastic flows. The inherent instability in the pure forward solver mandates our transition to a 10\% (2,000 points) sparse data assimilation strategy for all subsequent models.

\subsection{Varying Reynolds Number}

We swept the Reynolds number from 10 to 25 at a constant elasticity of $\lambda=0.1$. For these convective regimes, we deployed transfer learning alongside data assimilation, training the models for 5,000 Adam epochs followed by 10,000 L-BFGS iterations. 

\begin{figure*}[htbp]
    \centering
    \includegraphics[width=0.85\textwidth]{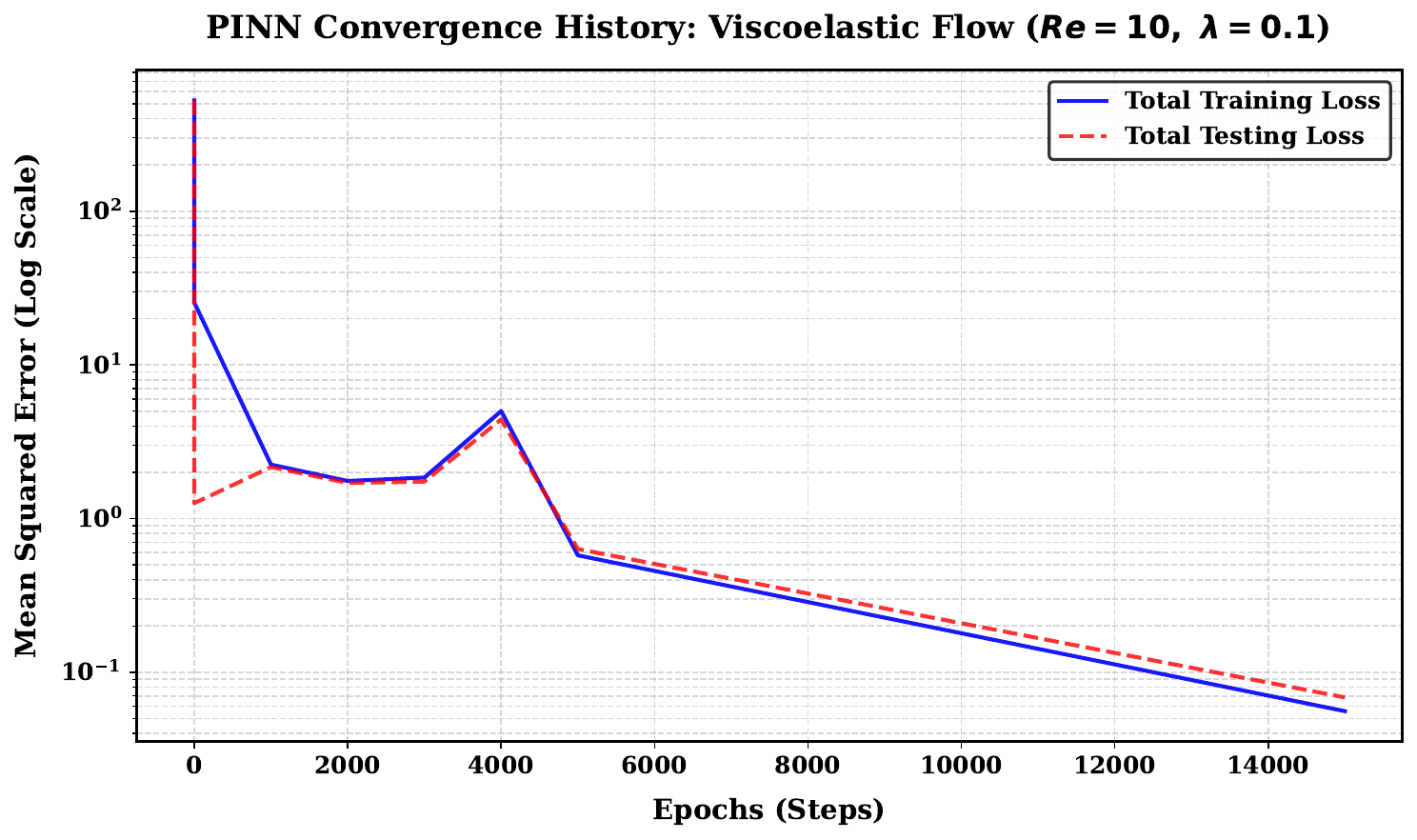}
    \caption{Convergence history for the data-assimilated PINN at $Re = 10$.}
    \label{fig:Re10_Loss}
\end{figure*}

\begin{figure*}[htbp]
    \centering
    \includegraphics[width=1.08\textwidth]{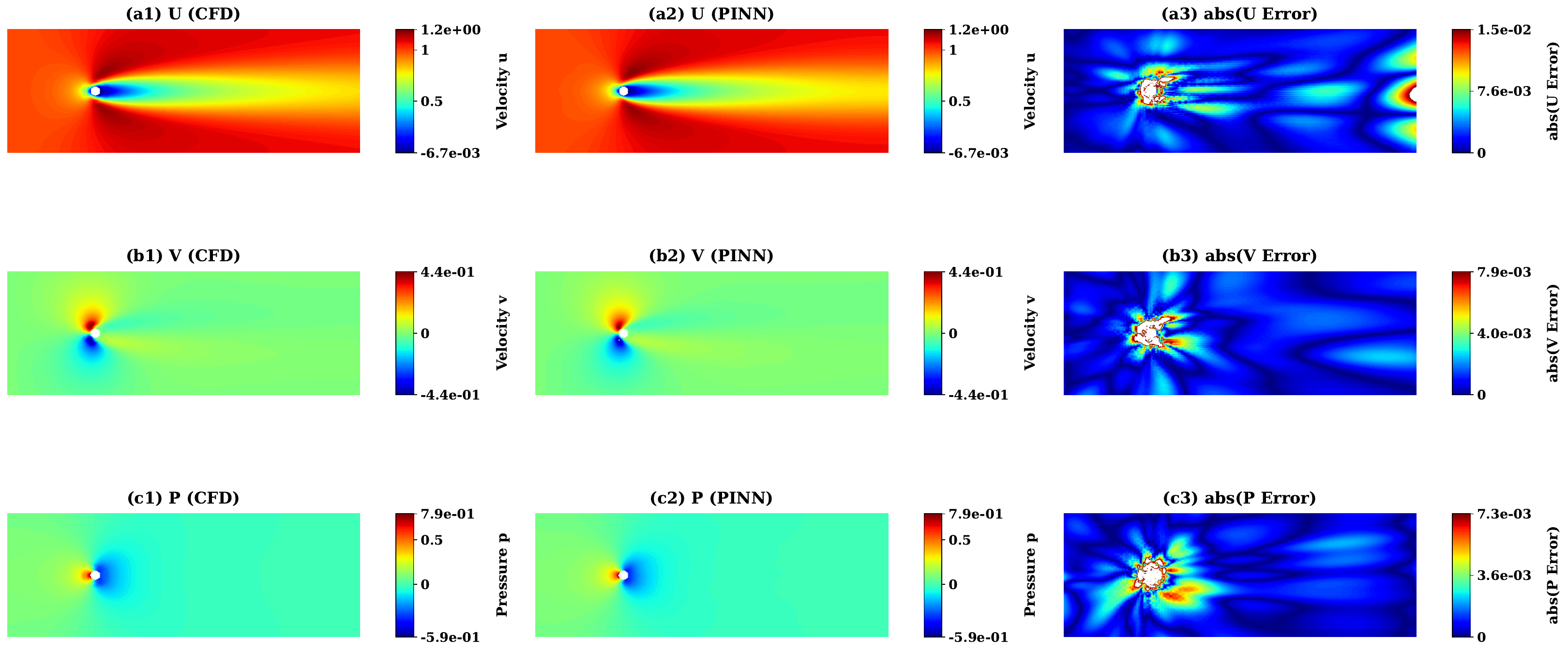} 
    \caption{For $\lambda = 0.1$ and $Re = 10$, a side-by-side comparison of the flow fields.}
    \label{fig:Re=10_cfd_vs_PINNs_comparison_master}
\end{figure*}

\begin{figure*}[htbp]
    \centering
    \includegraphics[width=0.85\textwidth]{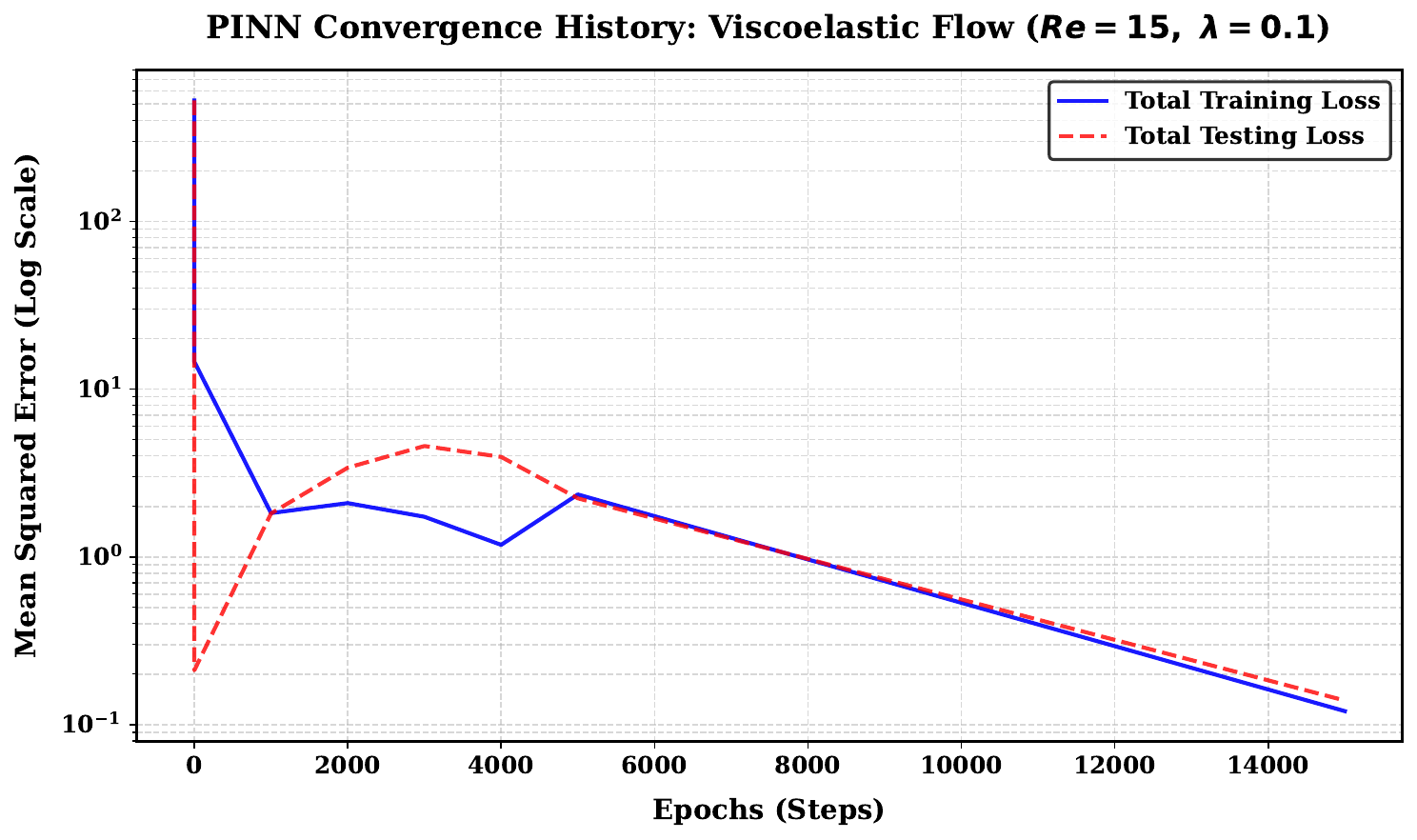}
    \caption{Convergence history for the data-assimilated PINN at $Re = 15$.}
    \label{fig:Re15_Loss}
\end{figure*}

\begin{figure*}[htbp]
    \centering
    \includegraphics[width=1.08\textwidth]{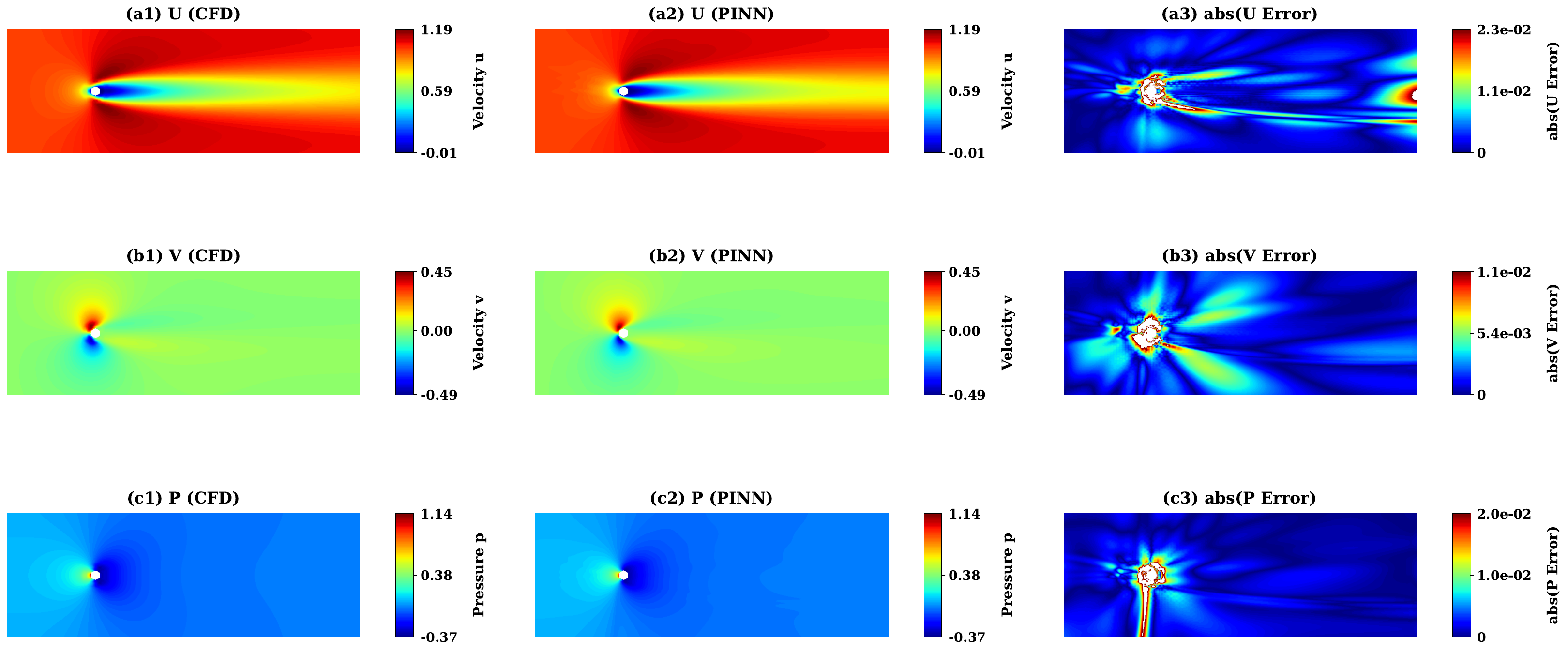}
    \caption{For $\lambda = 0.1$ and $Re = 15$, a side-by-side comparison of the flow fields.}
    \label{fig:Re=15_cfd_vs_PINNs_comparison_master}
\end{figure*}

\begin{figure*}[htbp]
    \centering
    \includegraphics[width=0.8\textwidth]{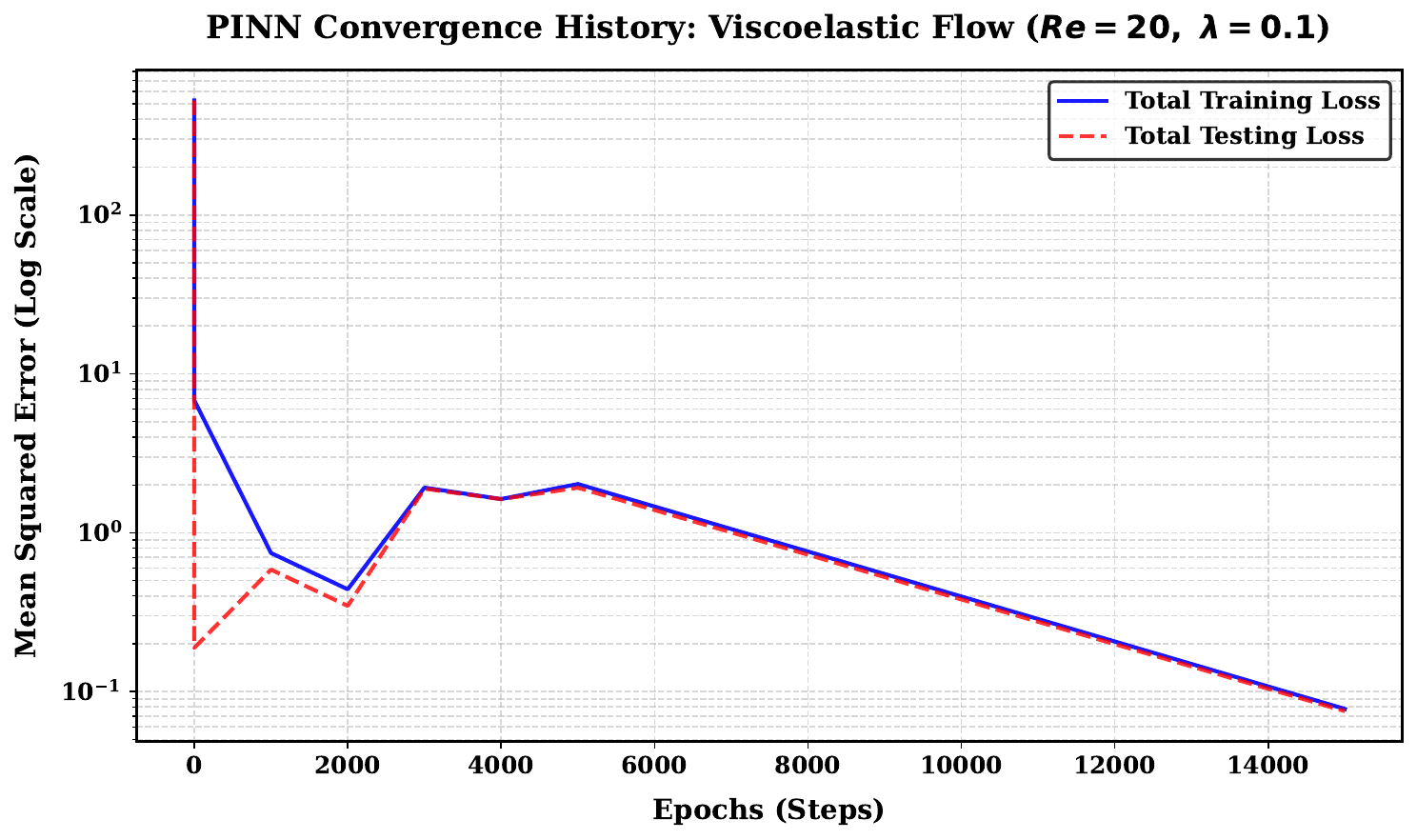}
    \caption{Convergence history for the data-assimilated PINN at $Re = 20$.}
    \label{fig:Re20_Loss}
\end{figure*}

\begin{figure*}[htbp]
    \centering
    \includegraphics[width=1.08\textwidth]{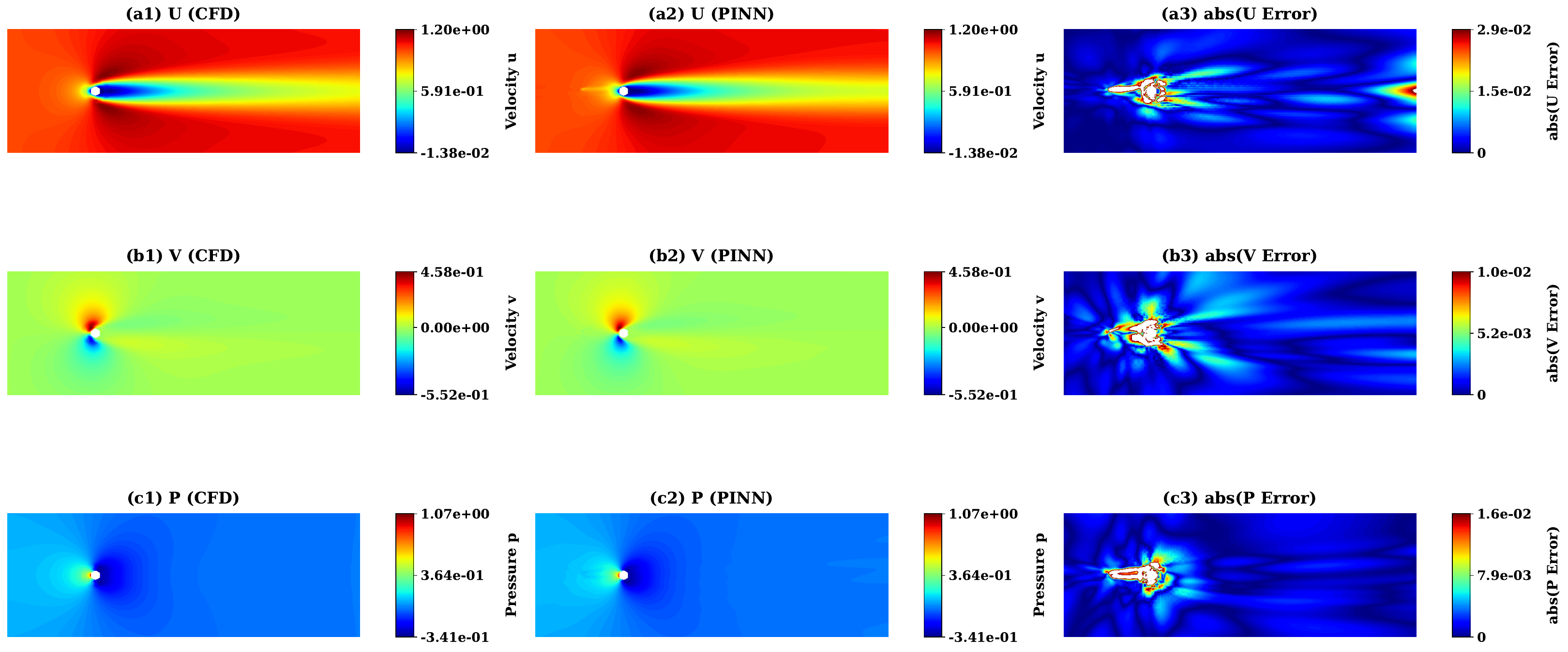}
    \caption{For $\lambda = 0.1$ and $Re = 20$, a side-by-side comparison of the flow fields.}
    \label{fig:Re=20_cfd_vs_PINNs_comparison_master}
\end{figure*}

\begin{figure*}[htbp]
    \centering
    \includegraphics[width=0.8\textwidth]{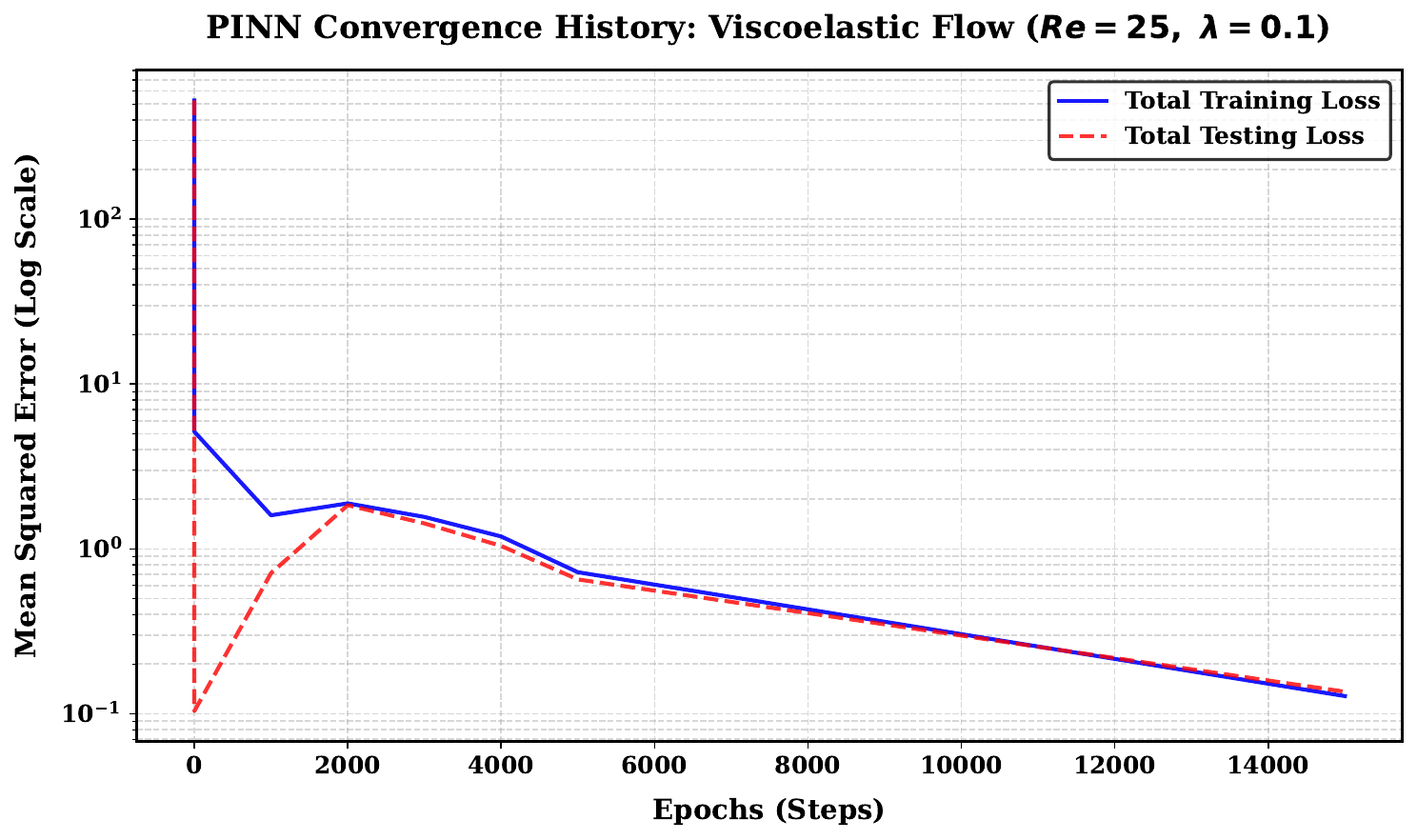}
    \caption{Convergence history for the data-assimilated PINN at $Re = 25$.}
    \label{fig:Re25_Loss}
\end{figure*}

\begin{figure*}[htbp]
    \centering
    \includegraphics[width=1.08\textwidth]{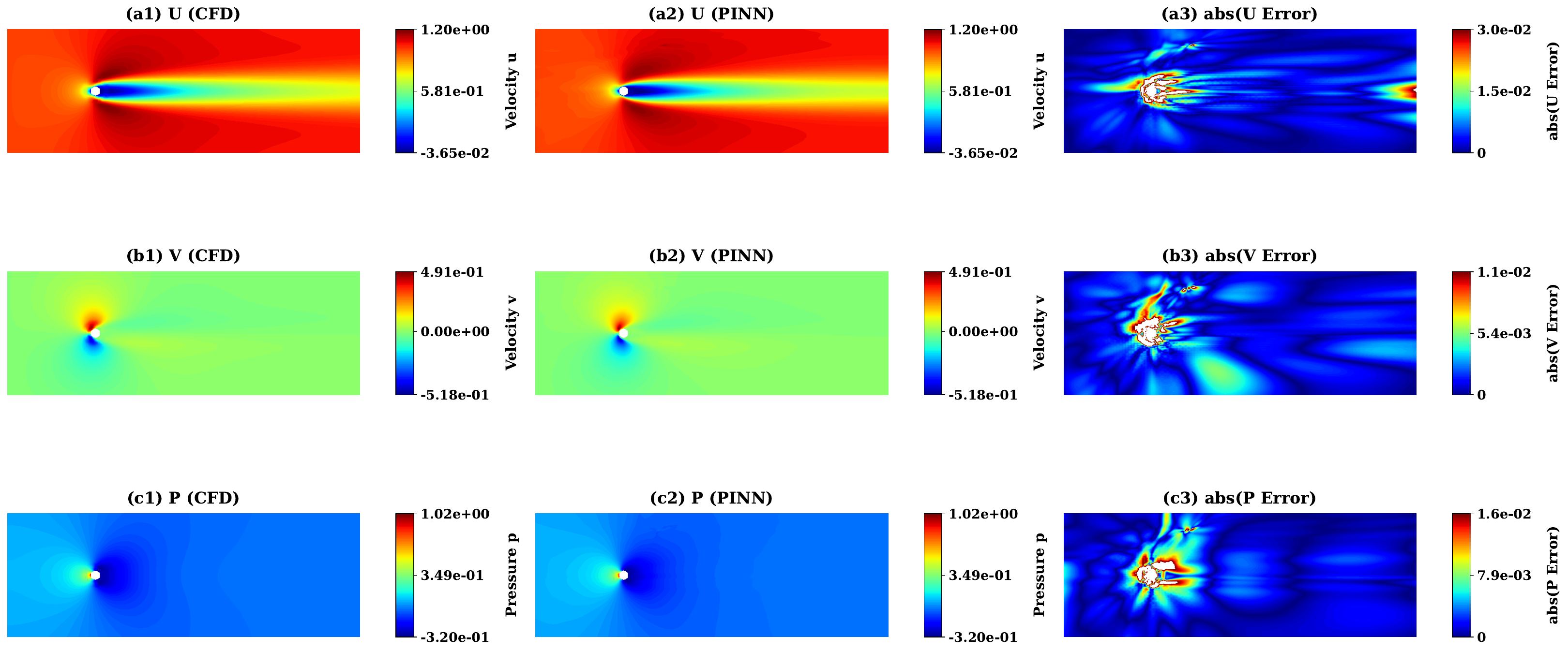}
    \caption{For $\lambda = 0.1$ and $Re = 25$, a side-by-side comparison of the flow fields.}
    \label{fig:Re=25_cfd_vs_PINNs_comparison_master}
\end{figure*}

When the Reynolds number is increased, the fluid's inertia shifts and the downstream wake elongates. However, the basic elastic behavior governed by the Weissenberg number remains largely unchanged. Since the underlying stress topology remains similar across these states, transfer learning becomes particularly effective. Restoring the converged weights from the $Re=5$ state acts as a computational foundation. It safely accelerates convergence for higher convective flows. The data-assimilated framework prevents the L-BFGS algorithm from exploiting interstitial gradient loopholes, keeping the testing and training error trajectories almost tightly aligned. In Fig.\ref{fig:Re10_Loss}, the convergence history at $Re=10$ shows both losses dropping steadily together below $10^{-1}$, which directly implies spatial accuracy. Looking at the flow fields in Fig.\ref{fig:Re=10_cfd_vs_PINNs_comparison_master}, the maximum absolute errors are heavily suppressed, reaching only $1.5\text{e-}02$ for the longitudinal velocity ($u$) and $7.3\text{e-}03$ for pressure ($p$). Scaling the inertia to $Re=15$ maintains this exact stability. The testing and training trajectories remain locked together during the L-BFGS optimization phase in Fig.\ref{fig:Re15_Loss}. Consequently, the error profiles in Fig.\ref{fig:Re=15_cfd_vs_PINNs_comparison_master} stay tightly bounded, with the absolute error magnitudes hovering around $10^{-2}$ across the domain. At $Re=20$, the paired loss trajectories drop smoothly toward $10^{-1}$ (Fig.\ref{fig:Re20_Loss}), restricting maximum absolute velocity errors in Fig.\ref{fig:Re=20_cfd_vs_PINNs_comparison_master} to just $2.9\text{e-}02$. Even at our highest Reynolds number case of $Re=25$, this locked loss descent continues (Fig.\ref{fig:Re25_Loss}). Despite the elongated wake, Fig.\ref{fig:Re=25_cfd_vs_PINNs_comparison_master} confirms maximum domain deviations never exceed $3.0\text{e-}02$ for velocity and $1.6\text{e-}02$ for pressure. 

\begin{figure*}[htbp]
    \centering
    \begin{minipage}{1\columnwidth}
        \centering
        \includegraphics[width=\linewidth]{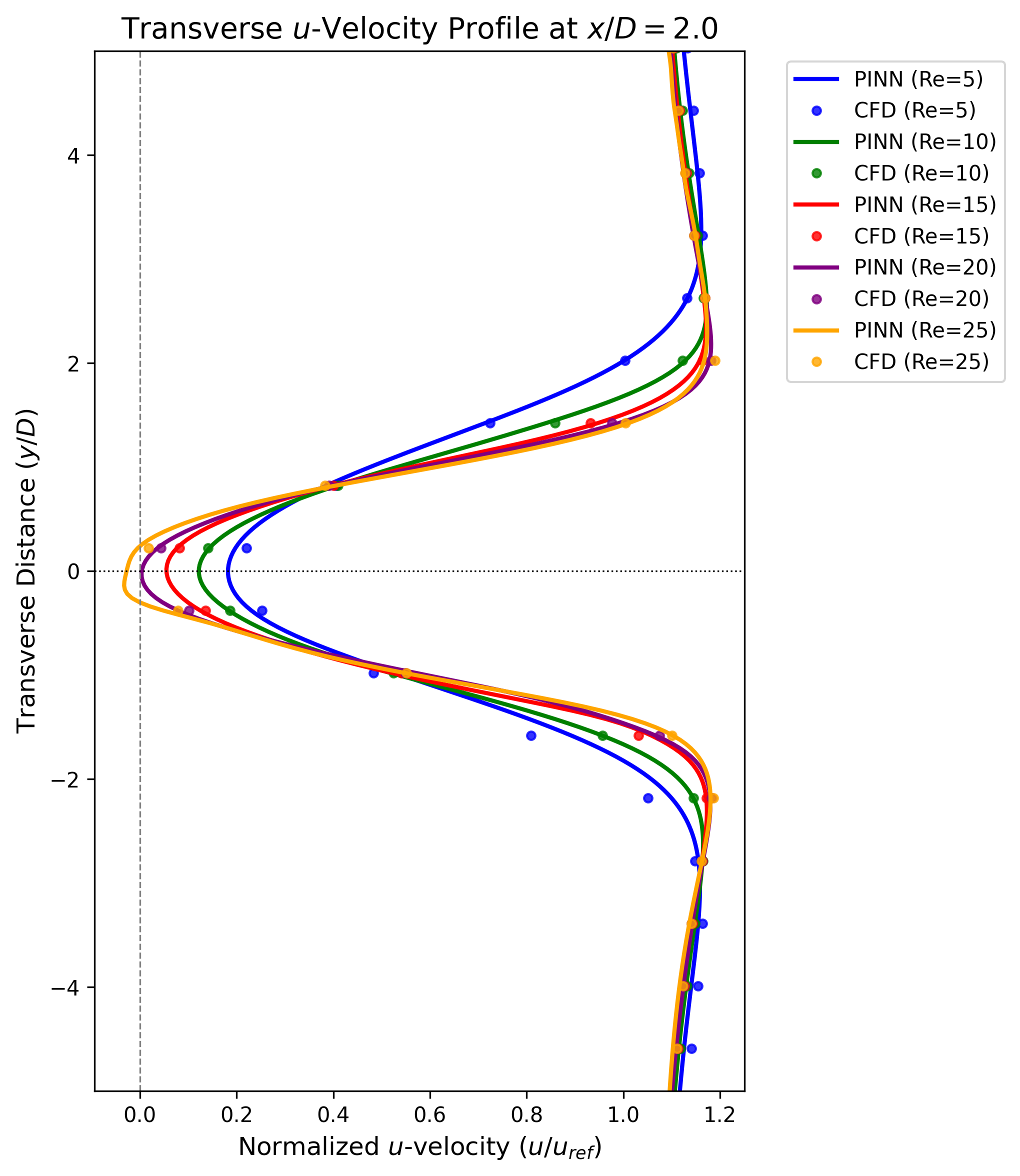}
        \caption{Transverse $u$ velocity profile at $x=2.0D$.}
        \label{fig:Linear_u_plot}
    \end{minipage}
    \hfill
    \begin{minipage}{1\columnwidth}
        \centering
        \includegraphics[width=\linewidth]{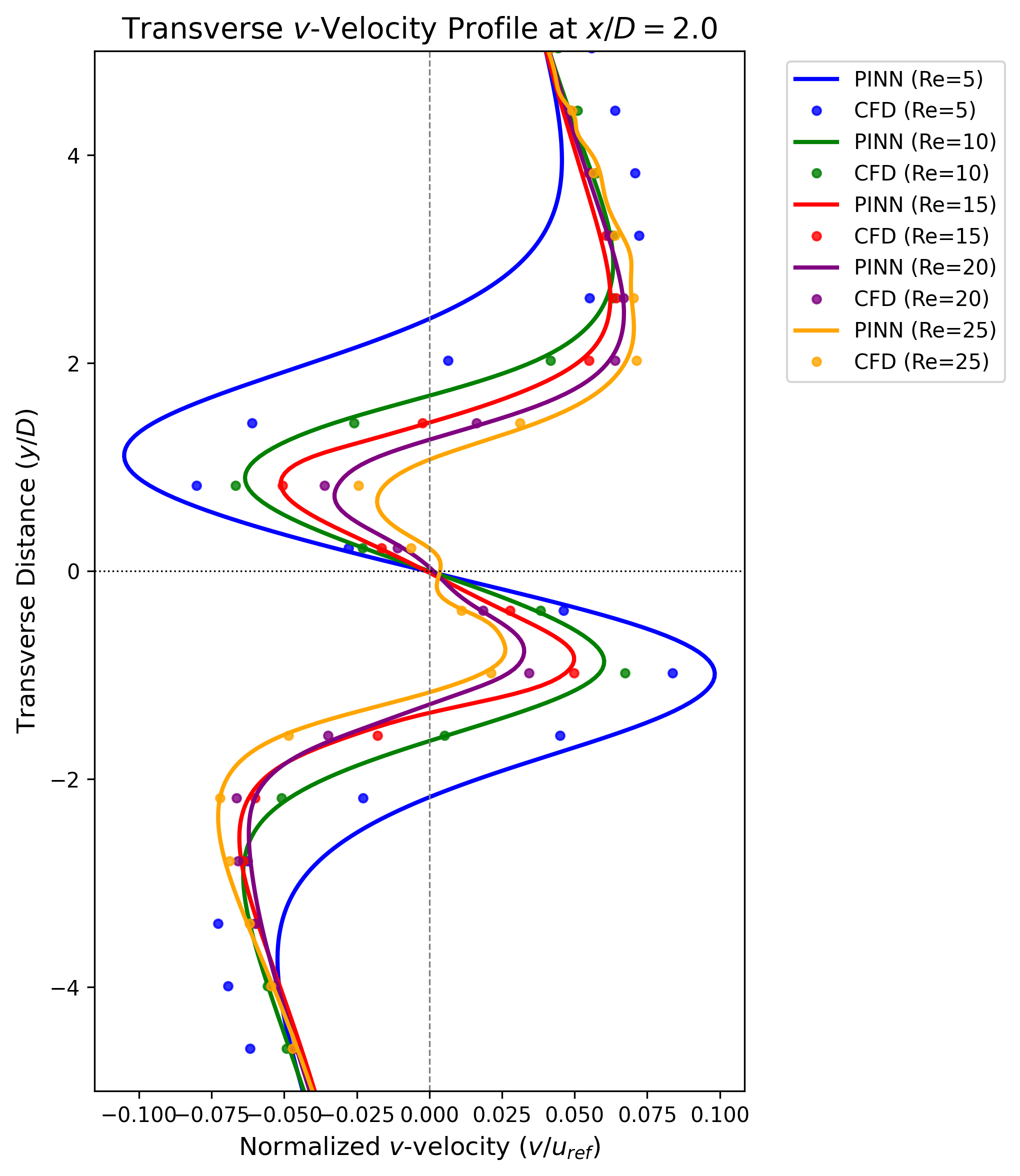}
        \caption{Transverse $v$ velocity profile at $x=2.0D$.}
        \label{fig:Linear_v_plot}
    \end{minipage}
\end{figure*}

While 2D contour maps validate the overall flow topology, we extracted 1D cross-sectional profiles to prove exact quantitative accuracy. Using the bulk inlet speed as our reference ($u_{ref} = 1.0$ m/s), Fig~.\ref{fig:Linear_u_plot} and Fig.~\ref{fig:Linear_v_plot} display the normalized transverse velocity profiles ($u/u_{ref}$ and $v/u_{ref}$) at the critical downstream location $x/D=2.0$.  

The solid lines representing the PINN predictions perfectly track the scattered OpenFOAM CFD data points. As the Reynolds number scales from 5 to 25, the wake thickens and the velocity deficit deepens. The PINN seamlessly captures this convective spreading. We observe no artificial numerical diffusion. The network perfectly matches the CFD scatter points even across the sharpest velocity gradients of the flow field.

\subsection{High Elasticity ($\lambda=0.3$ and $0.5$)}

 To rigorously challenge our data assimilation framework, we evaluated the flow at elevated relaxation times ($\lambda=0.3$ and $0.5$). We explicitly chose to train these high-elasticity cases from scratch. We disabled transfer learning entirely, relying strictly on a shortened, equalized training sequence of 5,000 Adam epochs and 5,000 L-BFGS iterations.

The physical justification for this decision is straightforward. Increasing $\lambda$ from 0.1 to 0.5 drastically escalates the Weissenberg number. The polymeric stress gradients grow exponentially near the cylinder walls, triggering the High Weissenberg Number Problem. While transfer learning serves as an excellent accelerator for parametric sweeps, utilizing it for these extreme elasticity cases could mask the true standalone capabilities of the baseline framework.

To eliminate any ambiguity and prove that convergence was not merely the result of a pre-trained computational crutch, we deliberately trained these highly elastic models from a completely random initialization. By achieving stable convergence using only the 2,000 data anchors alongside the Cholesky factorization, we demonstrate the inherent robustness of the data-assimilated framework in resolving stiff, high-Weissenberg fluid regimes.

\begin{figure*}[htbp]
    \centering
     \includegraphics[width=1.08\textwidth]{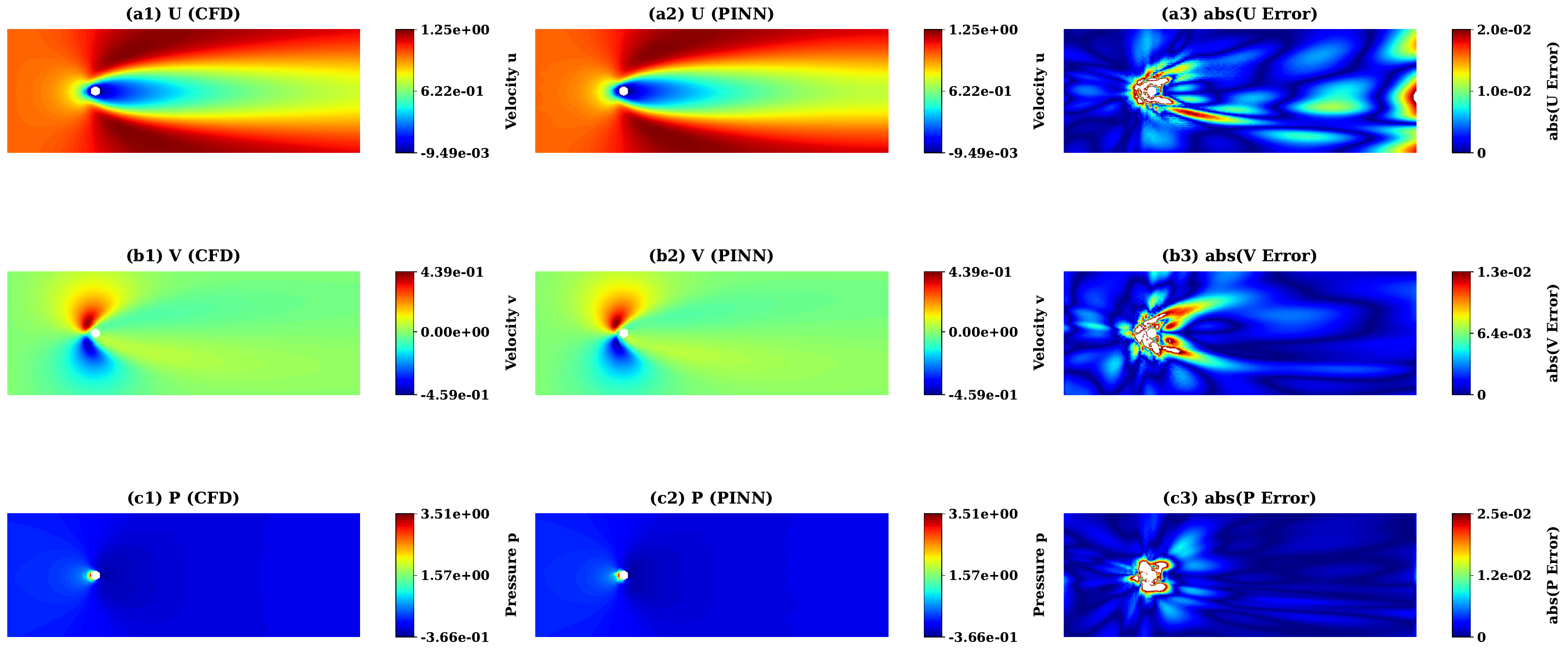}
    \caption{For $\lambda = 0.3$ and $Re = 5$, a side-by-side comparison of the flow fields.}
    \label{fig:Re=5_Lambda0.3_cfd_vs_PINNs_comparison_master}
\end{figure*}

\begin{figure*}[htbp]
    \centering
    \includegraphics[width=0.85\textwidth]{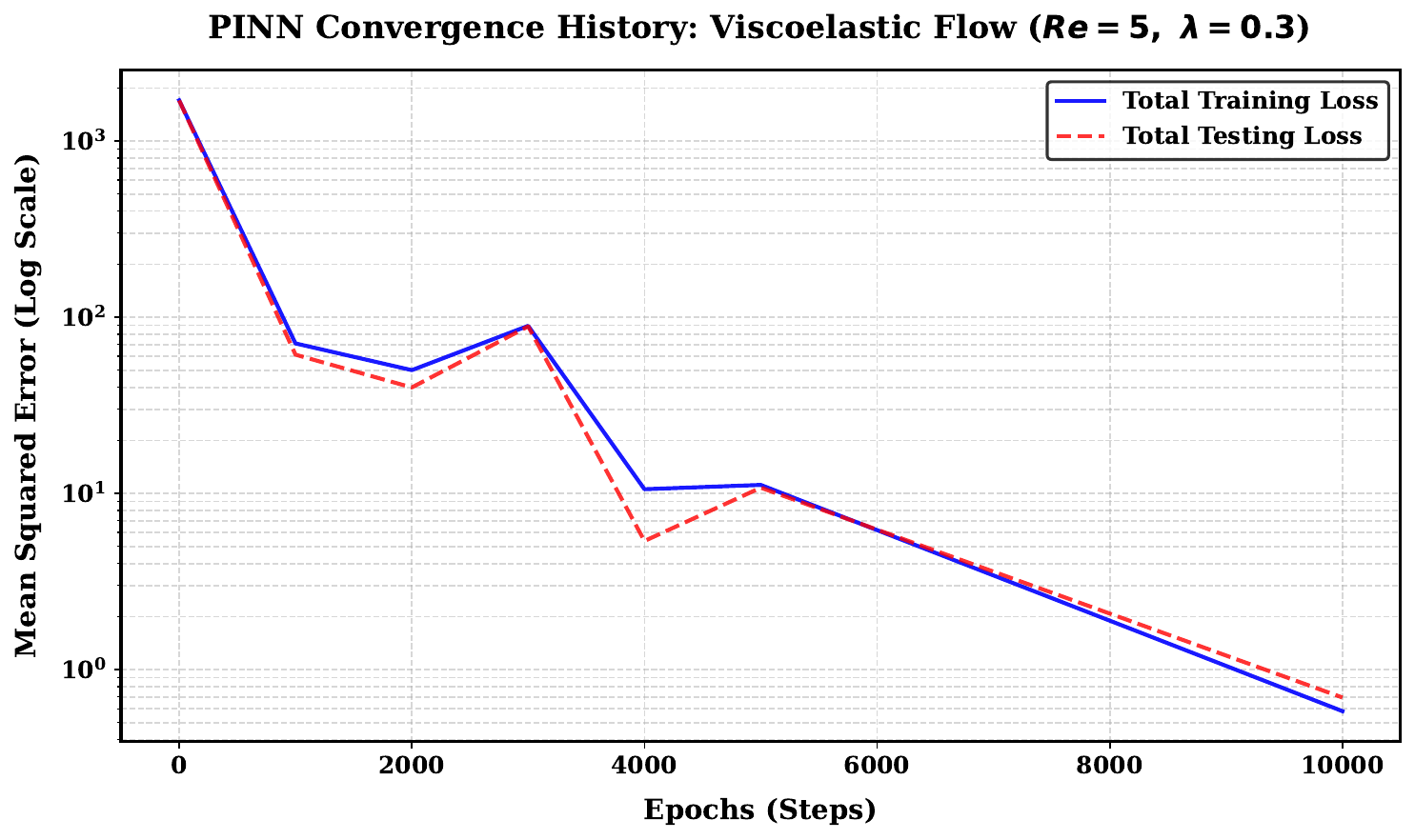}
    \caption{Convergence history for the data-assimilated PINN at $Re = 5$, $\lambda = 0.3$.}
    \label{fig:Re5_Lambda0.3_Loss}
\end{figure*}

\begin{figure*}[htbp]
    \centering
     \includegraphics[width=1.08\textwidth]{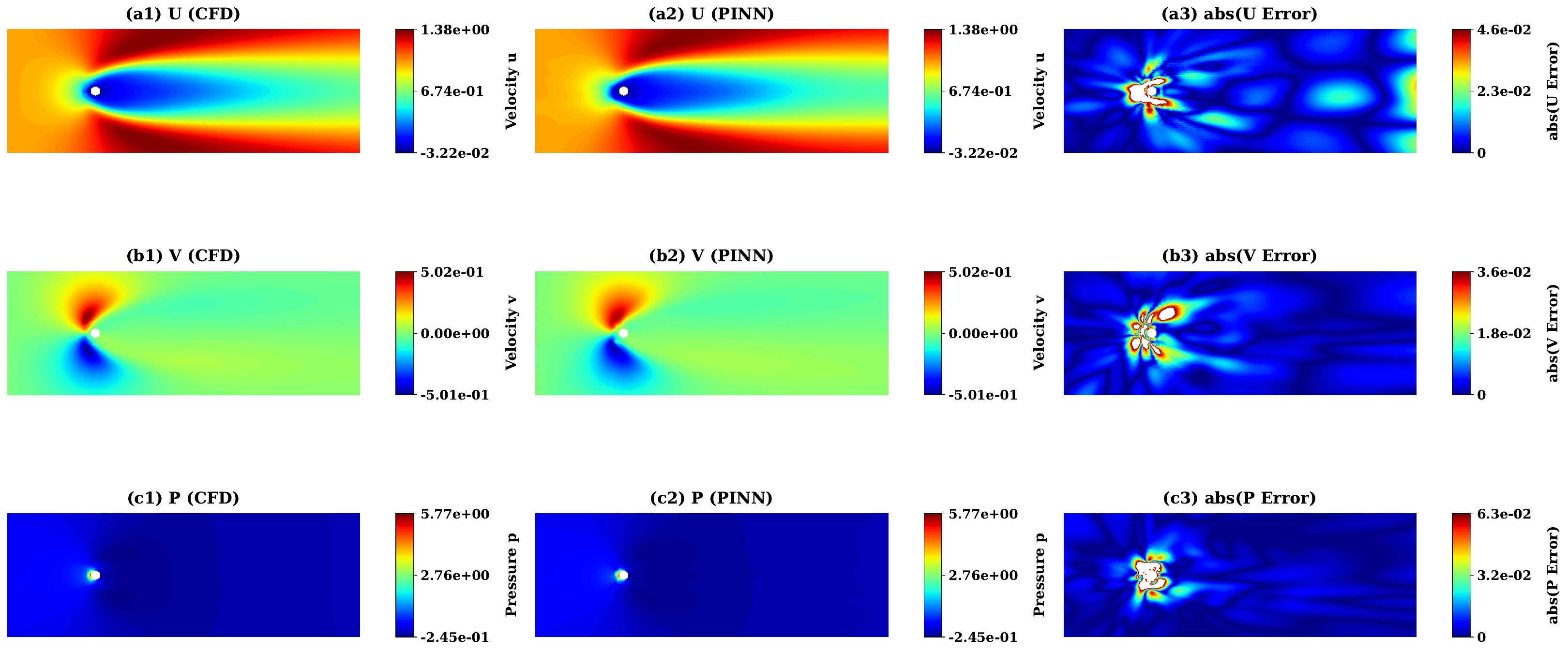}
    \caption{For $\lambda = 0.5$ and $Re = 5$, a side-by-side comparison of the flow fields.}
    \label{fig:Re=5_Lambda0.5_cfd_vs_PINNs_comparison_master}
\end{figure*}

\begin{figure*}[htbp]
    \centering
    \includegraphics[width=0.85\textwidth]{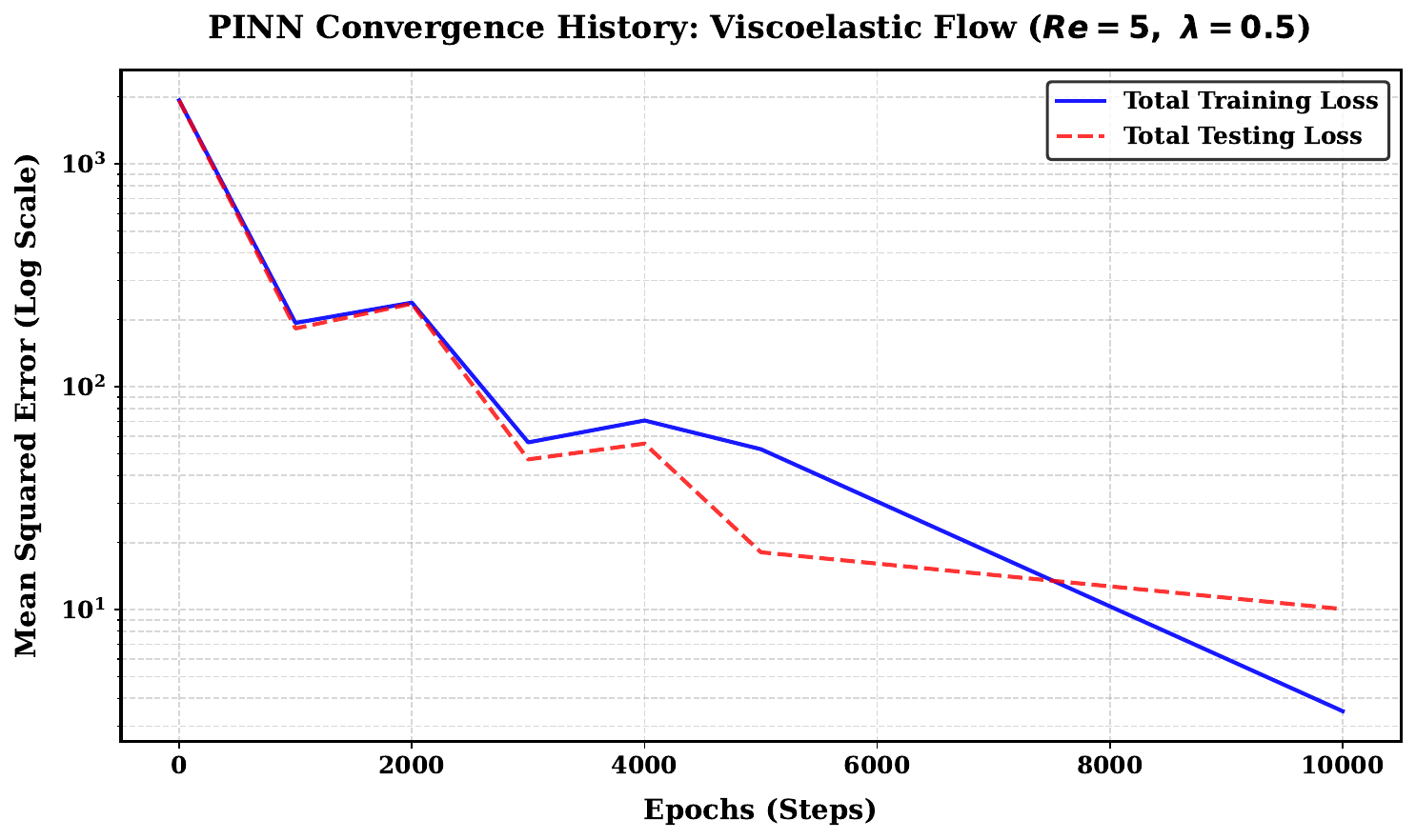}
    \caption{Convergence history for the data-assimilated PINN at $Re = 5$, $\lambda = 0.5$.}
    \label{fig:Re5_Lambda0.5_Loss}
\end{figure*}

The paired loss trajectories drive downward smoothly over the 10,000 epochs for both $\lambda=0.3$ (Fig.\ref{fig:Re5_Lambda0.3_Loss}) and $\lambda=0.5$ (Fig.\ref{fig:Re5_Lambda0.5_Loss}). As a result, the network achieves spatial accuracy entirely from scratch. For $\lambda=0.3$, Fig.\ref{fig:Re=5_Lambda0.3_cfd_vs_PINNs_comparison_master} shows maximum absolute velocity deviations that peak at only $2.0\text{e-}02$. Even at the extreme elasticity of $\lambda=0.5$, where polymeric stresses escalate massively, the absolute velocity error in Fig.\ref{fig:Re=5_Lambda0.5_cfd_vs_PINNs_comparison_master} remains heavily constrained to just $4.6\text{e-}02$ across the entire domain. 

\subsection{The Three-Cylinder Array}

We conclude our analysis by transitioning from isolated geometric primitives to a complex multi-body array. Modeling the narrow $0.5D$ gaps between the three staggered cylinders in OpenFOAM requires massive, computationally expensive mesh refinement to prevent truncation errors. Our PINN circumvents this entirely through Constructive Solid Geometry (CSG), completely bypassing grid generation.

We tested both Newtonian and viscoelastic configurations within this domain. The Newtonian network required a sparser domain of 12,000 points, solving only for the three primitive variables ($u, v, p$). In contrast, the viscoelastic Oldroyd-B network required a denser distribution of 20,000 points and six output variables ($u, v, p, l_{11}, l_{21}, l_{22}$) to resolve the Cholesky-decomposed stresses. 

\begin{figure*}[htbp]
    \centering
    \includegraphics[width=\textwidth]{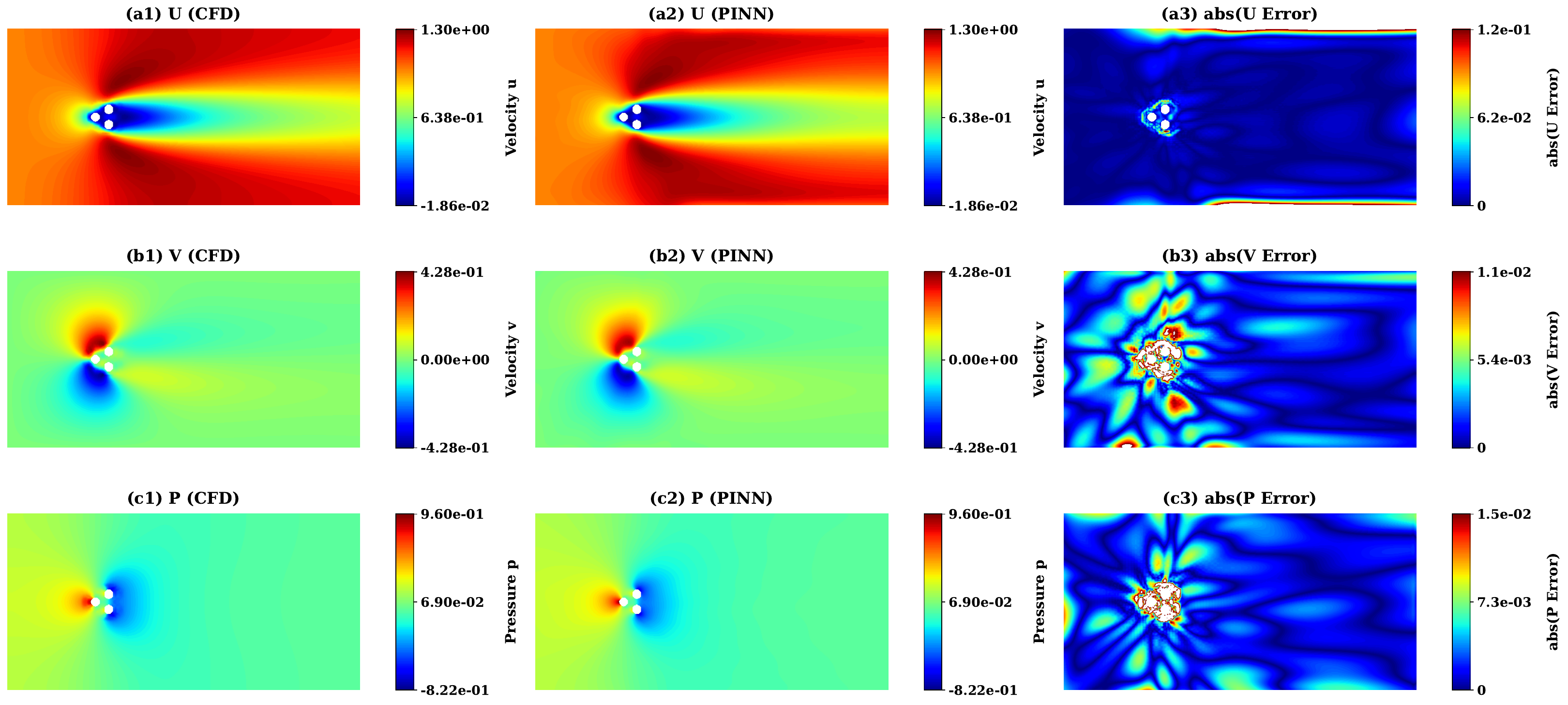}
    \caption{Flow field comparison and absolute localized error profiles for the newtonian fluid ($Re = 5$) interacting within the three-cylinder obstruction domain.}
    \label{fig:Newtonian_3cyl_master}
\end{figure*}

\begin{figure*}[htbp]
    \centering
    \includegraphics[width=\textwidth]{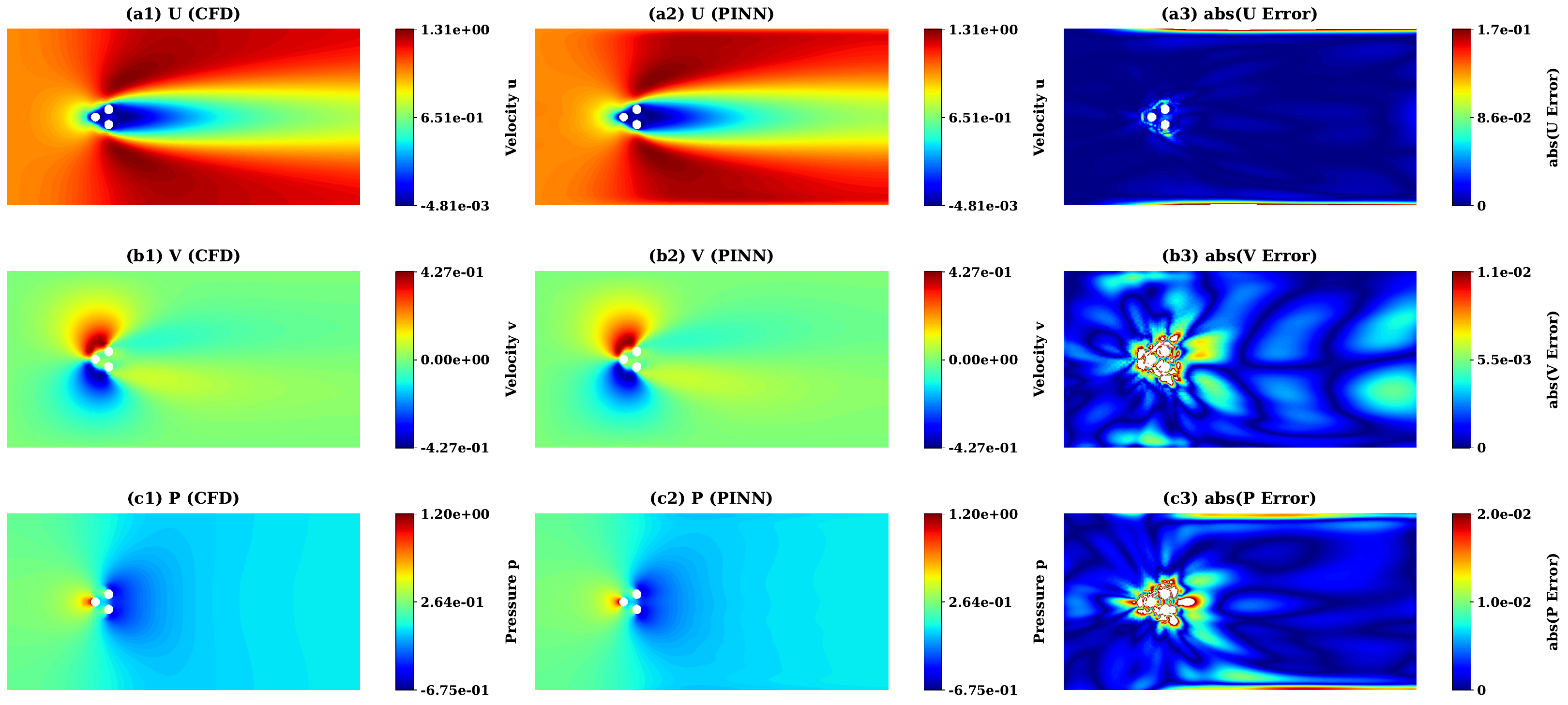}
    \caption{Flow field comparison and absolute localized error profiles for the viscoelastic Oldroyd-B fluid ($\lambda = 0.1$, $Re = 5$) interacting within the three-cylinder obstruction domain.}
    \label{fig:Visco_3cyl_master}
\end{figure*}

\begin{figure*}[htbp]
    \centering
    \includegraphics[width=0.8\textwidth]{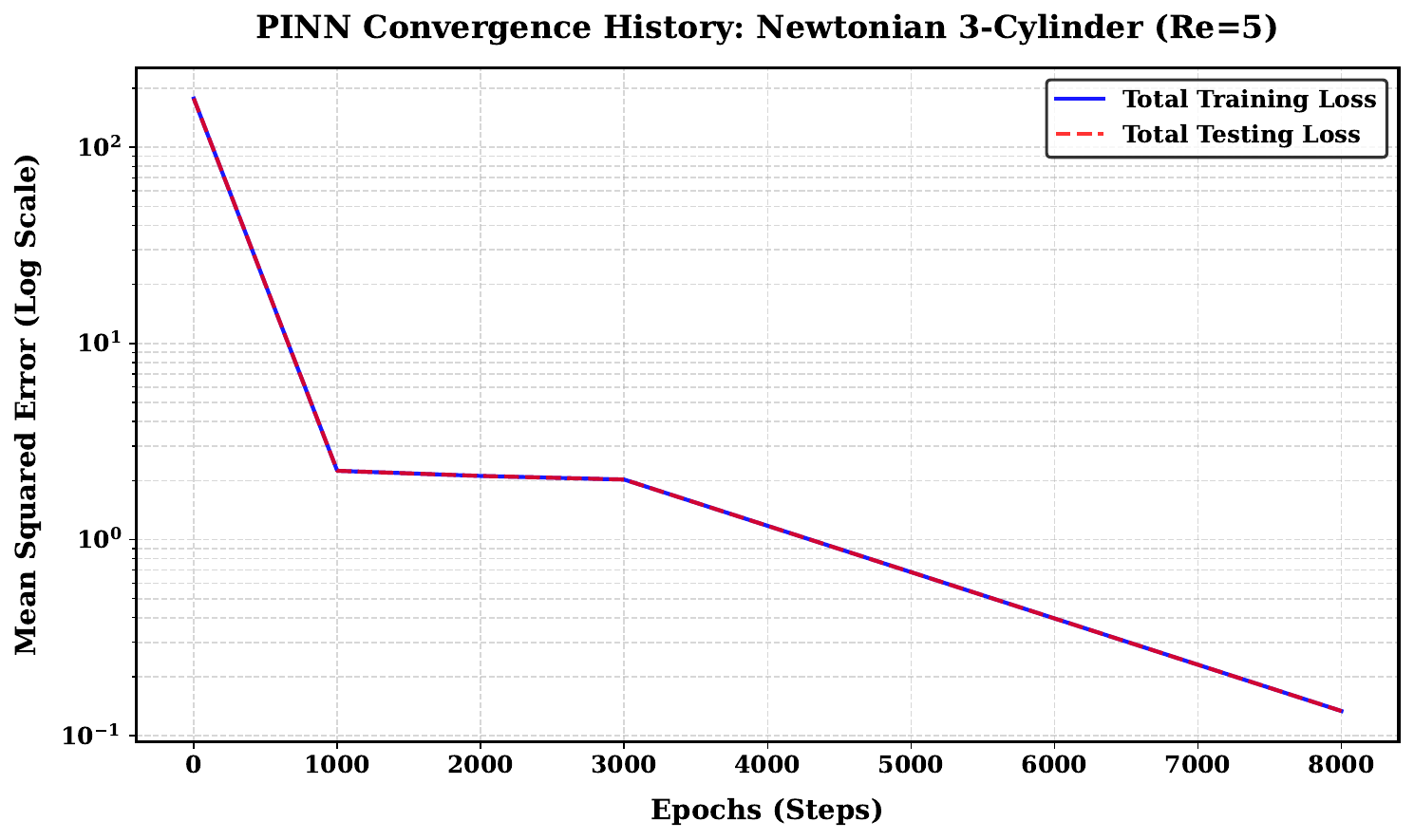}
    \caption{Convergence history for the data-assimilated Newtonian flow over the three-cylinder array ($Re=5$).}
    \label{fig:Newtonian_3Cyl_Loss}
\end{figure*}

\begin{figure*}[htbp]
    \centering
    \includegraphics[width=0.8\textwidth]{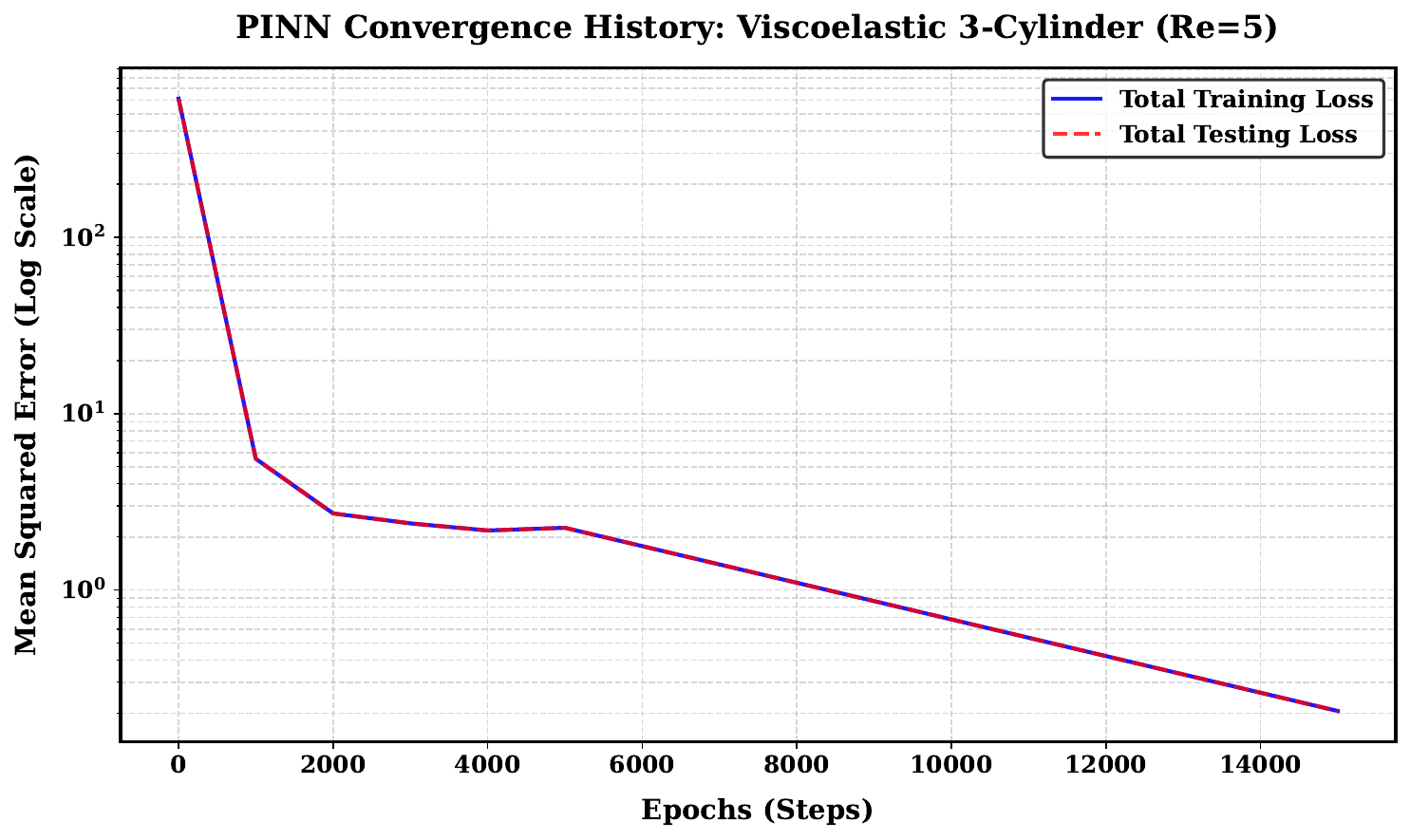}
    \caption{Convergence history for the viscoelastic Oldroyd-B configuration ($\lambda=0.1$, $Re=5$).}
    \label{fig:Visco_3Cyl_Loss}
\end{figure*}

Despite the extreme complexity of the overlapping elastic wakes and intense interstitial acceleration zones, the network successfully minimized the physics residuals. We maintained the exact same data assimilation strategy, utilizing 3,000 sparse anchors for both fluids. The paired loss trajectories drop smoothly and stabilize for both the Newtonian (Fig.\ref{fig:Newtonian_3Cyl_Loss}) and viscoelastic (Fig.\ref{fig:Visco_3Cyl_Loss}) configurations. For the Newtonian fluid, Fig.\ref{fig:Newtonian_3cyl_master} confirms the maximum absolute velocity error remains tightly bounded, peaking at $1.2\text{e-}01$. Even with the added mathematical stiffness of the conformation tensor, the viscoelastic network accurately resolves the complex flow topology, limiting the maximum velocity deviation in Fig.\ref{fig:Visco_3cyl_master} to just $1.7\text{e-}01$. This successful convergence across interacting solid boundaries definitively proves the universal geometric scalability of the data-assimilated Cholesky-PINN approach.

\section{Conclusion}

The present study establishes a robust hybrid Physics-Informed Neural Network (PINN) framework to resolve the complex viscoelastic dynamics of Oldroyd-B fluids. During our early testing, we found purely physical networks struggle with the severe mathematical stiffness of the conformation transport equations. As a result, they inevitably suffer from collocation point overfitting. We integrated a Cholesky tensor decomposition alongside a sparse data assimilation strategy to overcome this fundamental limitation. Through our work, we successfully overcome the High Weissenberg Number Problem and stabilized the steep boundary velocity gradients by anchoring the network with a small fraction of computational data (just $10\%$ of the domain). 

Our framework show exceptional physical adaptability across a wide variety of flow conditions. When dealing with higher convective inertia ($Re = 10$ to $25$), we paired our sparse data assimilation with transfer learning. Reusing prior converged states drastically cut down the optimization time. To push the limits on the highly elastic flows ($\lambda = 0.3$ and $0.5$), we dropped the transfer learning entirely and relied solely on data assimilation; the architecture still converged perfectly from a completely random initialization. Capturing extreme polymeric stress topologies does not actually require pre-trained weights. Passing this deliberate stress test proved the algorithm can stand completely on its own.

Constructive Solid Geometry (CSG) enabled a seamless transition to a complex three-cylinder array. This geometric scaling completely bypassed the severe mesh refinement bottlenecks traditionally plaguing finite volume solvers in narrow-gap configurations. Combining sparse data anchoring with Cholesky-factorized physics provides a highly stable, mesh-free, and computationally efficient surrogate for analyzing intricate non-Newtonian fluid interactions.

\section*{\textbf{Data and code availability}}
Data from this study and the computer scripts can be obtained from the authors upon 
reasonable request.
  
\section*{\textbf{Conflicts of Interest}}
No conflicts of interest, financial or otherwise, are declared by the authors.

\section*{Author Contributions} 
 MS, and AG planned the research;  MS carried out the calculations and analysed the numerical data; MS prepared the tables, figures, and the draft of the manuscript; MS and AG then revised the manuscript in detail and approved the final version.

\section*{\textbf{Acknowledgments}}
The authors would like to thank the Central Computing Resources at MANIT Bhopal, India, for providing computational support.
MS would like to acknowledge the UGC National Scholarship for Post Graduate Studies for financial support.  This research was supported in part by the International Centre for Theoretical Sciences (ICTS) for participating in the program - Plasma Turbulence: New Challenges from Magnetic Confinement 2026 (code: ICTS/PTMC2026/08). 

\bibliographystyle{unsrt}
\bibliography{refs}
\end{document}